\documentclass[12pt]{article}
\usepackage[totalwidth=460truept,totalheight=650truept]{geometry}
\usepackage{latexsym,graphicx,amsfonts,amssymb,amsmath,xcolor}
\usepackage[hypertexnames=false,hidelinks]{hyperref}
\usepackage{fancyhdr}

\def\theequation{\arabic{section}.\arabic{equation}}

\renewcommand{\theequation}{\thesection.\arabic{equation}}
\global\arraycolsep=1truept
\renewcommand{\theequation}{\arabic{section}.\arabic{equation}}

\begin{document}

\bigskip \phantom{C}

\vskip1truecm

\begin{center}
{\huge \textbf{Quantum Gravity}}

\vskip.5truecm

{\huge \textbf{with Purely Virtual Particles}}

\vskip.5truecm

{\huge \textbf{from Asymptotically Local}}

\vskip.5truecm

{\huge \textbf{Quantum Field Theory}}

\vskip1truecm

\textsl{Damiano Anselmi}

\vskip .1truecm

\textit{Dipartimento di Fisica \textquotedblleft Enrico Fermi", Universit%
\`{a} di Pisa}

\textit{Largo B. Pontecorvo 3, 56127 Pisa, Italy}

\textit{and INFN, Sezione di Pisa,}

\textit{Largo B. Pontecorvo 3, 56127 Pisa, Italy}

damiano.anselmi@unipi.it

\vskip1truecm

\textbf{Abstract}
\end{center}

We investigate the relationship between nonlocal and local quantum field
theories, and search for a viable notion of \textquotedblleft local
limit\textquotedblright\ to relate the unitary models. In Euclidean space it
is relatively easy to have nonlocal theories with well-behaved local limits.
In Minkowski spacetime, instead, singular behaviors are generically
expected. Relaxing some assumptions on the \textquotedblleft form
factors\textquotedblright\ considered in the literature, we identify a class
of models that have regular local limits in Minkowski spacetime. We call the
models \textquotedblleft asymptotically local\textquotedblright\ quantum
field theories (AL-QFTs) and show that their limits are theories with
physical and purely virtual particles (PVPs). In the bubble diagram, the
nonlocal deformation generates PVPs straightforwardly. In the triangle
diagram, it does so possibly up to multi-threshold corrections, which may be
adjusted by tuning the deformation itself. We also build an asymptotically
local deformation of quantum gravity with purely virtual particles. AL-QFT
can serve various purposes, such as suggesting innovative approaches to
off-shell physics, providing an alternative formulation for theories with
PVPs, or smoothing out nonanalytic behaviors. We discuss its inherent
arbitrariness and the implications for renormalizability.

\vfill\eject

\section{Introduction}

\label{intro}\setcounter{equation}{0}

Locality is a guiding principle of quantum field theory (QFT). At the same
time, quantum gravity (QG)\ challenges us to reconsider the principles that
worked successfully so far for the standard model of particle physics.
Adjusting the concept of locality is a possibility that must be taken into
consideration by physicists aiming to classify the viable theoretical
options.

In this paper we consider various possibilities in this regard. The main
goal is to work out a practically viable notion of \textquotedblleft local
limit\textquotedblright\ to relate local and nonlocal theories with
physically desirable properties, such as renormalizability and unitarity.
Besides deepening our knowledge on quantum fields, this investigation may
have interesting physical applications. Nonlocal deformations of the
standard model, for example, may be suitable to describe off-shell physics.
To our knowledge, the relationship between nonlocal and local theories has
not been investigated, so far, to the extent we aim to explore in this paper.

Various classes of nonlocal theories have been studied in the literature,
starting from Efimov \cite{efimovnl}. Krasnikov \cite{Krasnikov} had the
idea to remove the ghosts of local theories from the spectrum by means of
nonlocal deformations. Specifically, the unphysical poles of the free
propagators are replaced by certain form factors (entire functions with no
zeros) to ensure unitarity \cite{efimov,briscese}.

If the form factors are generic, they originate nonrenormalizable gauge and
gravity interactions. Concentrating on gravity, Kuz'min \cite{kuzmin} showed
that renormalizability can be obtained by choosing entire functions that
tend to polynomials at high energies. The parent local and the deformed
nonlocal theories are super-renormalizable. Precisely, the propagators and
vertices of the latter tend sufficiently fast to those of a
higher-derivative (HD) theory in the ultraviolet limit \cite{Lanza}.

Later, Tomboulis \cite{Tomboulis} revived this idea and applied it to gauge
theories. More recently, Modesto and others \cite{Modesto} elaborated it to
a greater extent and built a variety of finite models. Properties and
implications of more general form factors were studied in\ ref.s \cite%
{Koshelev}.

The class of super-renormalizable nonlocal theories just mentioned provides
an interesting arena for testing candidate notions of local limit. Naively,
we may expect that the limit\ is just the higher-derivative local theory
(HD-QFT) that describes the ultraviolet behavior. It turns out that the
relationship between locality and nonlocality is more subtle than that.

Let $\mathcal{M}_{\text{NL}}$ and $\mathcal{M}_{\text{L}}$ denote the
manifolds of nonlocal and local unitary theories (NL-QFTs and L-QFTs, from
now on), respectively. We look for families of models in $\mathcal{M}_{\text{%
NL}}$, parametrized by some variable $\lambda >0$, that become local when $%
\lambda \rightarrow \infty $. This way, the space $\mathcal{M}_{\text{L}}$
may be understood as a subspace of the boundary $\partial \mathcal{M}_{\text{%
NL}}$.

It is relatively easy to provide a meaningful notion of local limit in
Euclidean space. It is more difficult to extend the concept to Minkowski
spacetime. For example, in a class of treatable theories inspired by the
ones of the literature mentioned above, we find that the limit generates
severe singularities inside the correlation functions, such as integrals $%
\simeq \int \mathrm{d}^{4}p/|p^{2}-m^{2}|$. We are unable, at this very
moment, to say whether the nonlocal models of refs. \cite%
{kuzmin,Tomboulis,Modesto} admit proper local limits in Minkowski spacetime
or not.

This leads us to search for new, \textit{ad hoc} nonlocal models that do
have meaningful local limits in Minkowski spacetime when $\lambda $ tends to
infinity. We find that the form factors attached to the \textquotedblleft
propagators\ of nonpropagating fields\textquotedblright\ -- those which
eliminate the unwanted poles -- must have some zeros. Apart from that, they
are still entire functions. This generalization does not do too much harm,
since the associated fields can be integrated out with no loss. The
Lagrangian is singular, strictly speaking, but the action is regular.

Another direction to pursue to shed light on the matter is the relation
between the local limit $\lambda \rightarrow \infty $ and the continuation E 
$\rightarrow $ M from Euclidean space (E) to Minkowski spacetime (M). Do
these operations commute? If we take the local limit in E and then continue
from E to M, do we find the local limit in M?

The continuation E $\rightarrow $ M is not unique. In $\mathcal{M}_{\text{L}%
} $, for example, the analytic continuation of higher-derivative theories
gives ghosts (fields with kinetic terms multiplied by the wrong sign), while
the \textquotedblleft average continuation\textquotedblright\ \cite{AP}
gives theories with purely virtual particles (PVPs), i.e., particles that
are always off shell. By unitarity, the ghosts are unreachable from $%
\mathcal{M}_{\text{NL}}$, so the continuation E $\rightarrow $ M, which is
analytic for $\lambda <\infty $, cannot stay analytic when $\lambda
\rightarrow \infty $. The possibility that the target local limits contain
PVPs is compatible with this.

\bigskip

Summarizing, in this paper we build \textquotedblleft asymptotically
local\textquotedblright\ quantum field theories\ (AL-QFTs) -- models that
have regular local limits in Minkowski spacetime. Such limits contain purely
virtual particles in addition to physical particles (PVP-QFTs).

\bigskip

We recall two additional methods of introducing purely virtual particles,
besides the one already mentioned: one uses diagrammatic spectral optical
identities \cite{diagrammarMio} and the other one is based on non-time
ordered correlation functions of a special type \cite{PVP20}. Generically
speaking, PVPs are defined by tweaking the diagrammatics so as to eradicate
the ghosts, while remaining in the realm of local theories. That is to say,
the PVP and Krasnikov approaches share the same goals, in local and nonlocal
settings, respectively.\ We can also view AL-QFT as a fourth formulation of
theories with PVPs.

We also mention that, taking advantage of the PVP concept, it is possible to
build a\ local theory of quantum gravity \cite{LWgrav} that is unitary and
renormalizable at the same time. Its main prediction is a constrained window
($4/10000\lesssim r\lesssim 3/1000$) for the value of the tensor-to-scalar
ratio $r$ \cite{ABP} of primordial fluctuations, which essentially confirms
the prediction of the Starobinsky $R+R^{2}$ model \cite{staro} ($r\simeq
3/1000$) within less than an order of magnitude. Other predictions can be
derived as well (the running of scalar and tensor tilts, higher-order
corrections, etc. \cite{CMBrunning}), but require more effort to be tested,
in the realm of current or planned observations \cite{CMBStage4}. Among the
other things, in this paper we provide a nonlocal deformation of quantum
gravity with PVPs, which returns the theory of \cite{LWgrav} in the local
limit.

\bigskip

The local limit is straightforward at the tree level, where most of its
intriguing aspects are not visible. We push the investigation to the
simplest nontrivial diagrams, that it to say, the bubble and the triangle.
We show that in the case of the bubble diagram the nonlocal deformation
\textquotedblleft already knows\textquotedblright , so to speak, how to
generate PVPs, with no need of \textit{ad hoc} adjustments. In the cases of
triangle and more complicated diagrams, the local limit is \textquotedblleft
PVP ready\textquotedblright\ in the null and single-threshold sectors. The
multi-threshold sectors, instead, are more difficult to handle. Possibly,
the arbitrariness of the deformation must be invoked to fine-tune the limit
appropriately.

The weakness of nonlocal quantum field theory is that it lacks a fundamental
principle to select the nonpolynomial functions it is built upon, remove its
inherent arbitrariness and identify a unique candidate theory of nature.
However, we do not propose AL-QFT as a framework for fundamental theories.
We just claim that AL-QFT is interesting \textit{per se}, broadens our
knowledge of QFT, and provides useful tools in support of local QFT. For
example, as an alternative formulation of PVPs it may allow us to address
peculiar aspects of these \textquotedblleft non particles\textquotedblright\
(such as the violation of microcausality \cite{causalityQG} and the so
called \textquotedblleft peak uncertainty\textquotedblright\ \cite{peak}).

Another instance where a comparable degree of arbitrariness plagues (local)
QFT (without jeopardizing it\ as a framework for fundamental interactions)
is off-the-mass shell physics \cite{offshell}. There, the extra parameters
describe the environment where the phenomenon under observation is placed.
AL-QFT may provide an alternative way to describe these aspects. In this
respect, the lack of uniqueness of AL-QFT is not an issue.

Another way to settle the non uniqueness of AL-QFT is as follows. We have
remarked that nonlocal theories lack a selection principle for the form
factors. The local limit might serve as one. Then, the candidate theories to
describe nature from AL-QFT are nothing but their local limits\footnote{%
Strictly speaking, they are not contained in the manifold $\mathcal{M}_{%
\text{AL}}$ of asymptotically local theories, but in its border $\partial 
\mathcal{M}_{\text{AL}}$.}.

\bigskip

The paper is organized as follows.

\begin{enumerate}
\item In section \ref{AL} we recall the basic features of the nonlocal
theories considered in the literature and explain how we are going to
generalize them to have well-defined local limits in Minkowski spacetime.

\item In section \ref{localE} we study the local limit in Euclidean space,
which is relatively straightforward.

\item In section \ref{WickDirect} we describe the local limit in Minkowski
spacetime, including explicit calculations for the bubble and triangle
diagrams. We compare the behaviors of the nonlocal theories of the
liturature with the properties of the new models built in section \ref{AL}.

\item In section \ref{alqg} we formulate an asymptotically local deformation
of quantum gravity with purely virtual particles along the lines of section %
\ref{AL}.
\end{enumerate}

Section \ref{conclusions} contains the conclusions. In appendix \ref{approx}
we build the key functions that are needed for the form factors used in the
paper. In appendix \ref{app1} we recall the calculation of the bubble
diagram with PVPs. When necessary, the dimensional regularization \cite%
{dimreg} is used for explicit calculations, where $D=4-\varepsilon $ denotes
the continued dimension around dimension 4.

\section{The nonlocal theories we consider here}

\label{AL}\setcounter{equation}{0}

We want to study the local limits of nonlocal theories. There is no
guarantee that such a limit exists, in general, so we may have to construct 
\textit{ad hoc} models that admit one.

In this section we lay out the basic features of the classes of theories
under consideration. We include those mostly studied in the current
literature, as well as extensions that have well-behaved local limits.

We first work in Euclidean space and later switch to Minkowski spacetime by
means of the analytic continuation. The models we focus on have Euclidean
Lagrangians of the form 
\begin{equation}
\mathcal{L}_{\text{NL}}=\frac{\tau }{2}\phi (-p)Q(P(p))\phi (p)+\mathcal{O}%
(\phi ^{3}),  \label{nloc}
\end{equation}%
in momentum space, where $\phi $ denotes the fields, $Q(P)$ is a certain
function of a polynomial $P$ of the Euclidean momentum $p$, and $\tau =\pm 1$
is there so that we can describe nonlocal deformations of both physical
particles and ghosts.

We take 
\begin{equation}
P(p)=p^{2}+m^{2}  \label{Pp}
\end{equation}%
where $m$ is the \textquotedblleft mass\textquotedblright\ in the local
limit. Polynomials of higher-derivative theories, such as%
\begin{equation}
P_{\text{HD}}(p)=(p^{2}+m^{2})^{2}+M^{4},  \label{PHD}
\end{equation}%
are also interesting, but will not be treated in detail here.

For a while, we concentrate on the propagator and study the local limit
there. \ We assume that the vertices are local or their limits do not
invalidate the derivations we make. We discuss the vertices in detail in
subsection \ref{vertices}.

In this context, the key quantity that encodes the nonlocal model is the
function $Q(P)$. If we want a degree of freedom from $\phi $, we know how to
obtain it: $Q=P$. The challenge is to \textit{not} have a degree of freedom
from $\phi $. That is to say, $\phi $ should be purely virtual\footnote{%
We could call it \textquotedblleft nonlocal purely virtual
particle\textquotedblright\ (NL-PVP).}. To achieve this, we require that $%
\tau /Q(P)$ is entire in the complex $P$ plane, so that $Q(P)$ has no zeros.
In addition, we require that $Q(P)$ tends to $P$ in the local limit defined
below. This makes $\phi $ a natural candidate to become an ordinary PVP in
that limit.

The main difference with respect to the assumptions commonly adopted in the
literature \cite{kuzmin,Tomboulis,Modesto} is that the propagator $\tau /Q(P)
$ is not required to never vanish in the complex $P$ plane. Specifically, we
allow $Q(P(p))$ to have a singularity proportional to $(-p_{\text{M}%
}^{2}+M^{2})^{-1}$, where $p_{\text{M}}$ is the Minkowskian momentum and $M$
is some mass. The Minkowskian action is then defined by means of the Cauchy
principal value.

The Lagrangian (\ref{nloc}) can describe the $\phi $ subsector of a more
general theory, which may contain ordinary physical particles $\varphi $ as
well, as described by the extension 
\begin{equation}
\mathcal{L}_{\text{NL}}^{\prime }=\frac{\tau }{2}\phi (-p)Q(P(p))\phi (p)+%
\frac{1}{2}\varphi (-p)(p^{2}+m^{2})\varphi (p)+\mathcal{L}_{\text{int}%
}^{\prime },  \label{nlocca}
\end{equation}%
where $\mathcal{L}_{\text{int}}^{\prime }$ collects the interactions. We
focus our attention on $\phi $ here, since the propagators of the physical
particles $\varphi $ do not need to be nonlocally deformed.

\subsection{Local limit}

To begin with, we define the local limit as follows: we rescale $P$ and $Q$
by $\lambda $ and $\lambda ^{-1}$, respectively, where $\lambda $ is a
positive factor, and then let $\lambda $ tend to infinity. We assume that $Q$
tends to $P$ on the real axis: 
\begin{equation}
\lim_{\lambda \rightarrow +\infty }\lambda ^{-1}Q(\lambda P)=P,\qquad P\in 
\mathbb{R}.  \label{localass}
\end{equation}

The functions $Q$ we consider in this paper are%
\begin{equation}
Q(P)=h(P),\qquad Q(P)=\frac{h^{2}(P)}{P},  \label{Q}
\end{equation}%
where $h(P)$ is the entire function defined in formula (\ref{hz}), taken
from the current literature on nonlocal theories \cite%
{kuzmin,Tomboulis,Modesto}.

In appendix \ref{approx} we show that $h(P)$ tends to the absolute value, $%
P/h(P)$ tends to the sign function and $P/h^{2}(P)$ tends to the Cauchy
principal value in the local limit. The second option of (\ref{Q}) is the
one that allows us to build the asymptotically local models.

With choices like (\ref{Q}), the property (\ref{localass}) extends to a
double cone $\mathcal{C}$ around the real axis (see appendix \ref{approx}).

\subsection{Propagator}

The propagator of the Euclidean theories (\ref{nloc}) is%
\begin{equation}
G_{\text{nl}}(p)=\frac{\tau }{Q(P(p))},  \label{Gnl}
\end{equation}%
where the polynomial $P(p)$ is given by (\ref{Pp}), the function $Q$ is
given in (\ref{Q}) and $h$ given by (\ref{hz}). In the local limit, both
choices (\ref{Q}) give $\tau /P$ in Euclidean space. As we are going to
show, the two options lead to crucially different results in Minkowski
spacetime.

Associated with $G_{\text{nl}}(p)$, we have a double cone $\mathcal{C}$
where the expansion (\ref{expansion}) applies.

\subsection{Loop integrals}

Now we study general properties of the Feynman diagrams of nonlocal
theories. Let $p,p_{i}$ denote the internal momenta and $k,k_{a}$ the
external ones. The loop integral associated with a diagram $F$ is the
integral of a product of propagators $G_{\text{nl}}(p_{i})$ times a product
of functions $V(p_{i},k_{a})$ originated by the vertices. The latter are
also entire functions, as we explain in subsection \ref{vertices}, so we
concentrate on the propagators.

For example, the bubble diagram with circulating $\phi $ fields has the form%
\begin{equation}
\mathcal{B}(k)=\int \frac{\mathrm{d}^{D}p}{(2\pi )^{D}}%
V_{1}(p,k)V_{2}(p,k)G_{\text{nl}}(p)G_{\text{nl}}(p-k).  \label{bub}
\end{equation}

\bigskip

The integrals must be defined in Euclidean space. This means that both $%
p_{i} $ and $k_{a}$ are Euclidean momenta. Only the external momenta $k_{a}$
are later analytically continued to Minkowski spacetime. We denote their
Minkowski versions by $k_{\text{M}}$.

It is easy to show that the integrals are convergent for Euclidean $k_{a}$,
in the sense of the dimensional regularization. Consider, for example, the
expression (\ref{bub}) for $\mathcal{B}(k)$. For Euclidean $p$ and $k$, the
arguments $P(p)$ and $P(p-k)$ are located inside the cones $\mathcal{C}$ and 
$\mathcal{C}^{\prime }$ associated with the propagators $G_{\text{nl}}(p)$
and $G_{\text{nl}}(p-k)$. Formula (\ref{expansion}) then ensures that (\ref%
{bub}) tends to the same integral with $G_{\text{nl}}(p)\rightarrow P(p)$
and $G_{\text{nl}}(p-k)\rightarrow P(p-k)$, which is indeed convergent in
the sense of the dimensional regularization (assuming that the vertices do
not invalidate the argument, see \ref{vertices}). What this means is that
there exists an open set $\Omega $ of the complex $D$ plane where the
integral is convergent in the standard sense, or that the integral can be
split into a finite sum of integrals that admit convergence domains $\Omega
_{i}$ in the standard sense \cite{renormalization}.

Now we prove that the loop integrals are convergent, again in the sense of
the dimensional regularization, for every \textit{complex} external momenta $%
k_{a}$. Actually, they are entire functions of $k_{a}$, so it is possible to
replace the external momenta $k$ with their Minkowski versions $k_{\text{M}}$
directly inside the integrals.

It is important to stress that the integrated momenta $p_{i}$ remain
Euclidean till the very end. The reason is that it is not convenient to make
a Wick rotation on them. When we close integration paths by including arcs
at infinity, we cross regions where the functions $G_{\text{nl}}(p)$ and $G_{%
\text{nl}}(p-k)$ behave in ways that are hard to control.

Consider (\ref{bub}) again.\ When $k$ is deformed to complex values, the
cone $\mathcal{C}^{\prime }$, which is equal to $\mathcal{C}$ translated by $%
k$, is no longer centered along the Euclidean domain, but somewhere else,
depending on the imaginary part of $k$: 
\begin{equation*}
P(p-k)=(p-k)^{2}+m^{2}=\rho \mathrm{e}^{i\theta },\qquad \theta =\arctan 
\frac{\text{Im}[k_{4}(k_{4}-2p_{4})]}{\text{Re}[(p-k)^{2}+m^{2}]}.
\end{equation*}%
The phase $\theta $ of $P(p-k)$ tends to zero when the integrated momentum $p
$ tends to infinity, in any direction. This implies that the argument of $G_{%
\text{nl}}(p-k)$ falls off fast enough inside $\mathcal{C}^{\prime }$ for
sufficiently large $p$, which makes the integral (\ref{BM}) convergent.

Together with the fact that the integrand is regular everywhere (both the
vertices and the propagators being entire functions, see section \ref%
{vertices}), this argument proves that the function $\mathcal{B}(k)$ is
entire. Then we can perform the analytic continuation E $\rightarrow $\ M by
replacing $k=(k_{4},\mathbf{k})$ with its Minkowskian version $k_{\text{M}%
}=(-ik^{0},\mathbf{k})$ directly inside. The Minkowskian bubble diagram is
thus%
\begin{equation}
\mathcal{B}_{\text{M}}(k_{\text{M}})=i\int \frac{\mathrm{d}^{D}p}{(2\pi )^{D}%
}V_{1}(p,k_{\text{M}})V_{2}(p,k_{\text{M}})G_{\text{nl}}(p)G_{\text{nl}%
}(p-k_{\text{M}}),  \label{BM}
\end{equation}%
the factor $i$ being due to the fact that $p$ remains Euclidean.

The property just proved extends to more complex loop integrals, which also
define entire functions of $k_{a}$. Note that it does not apply, on the
contrary, to local quantum field theory. There, the entire functions $G_{%
\text{nl}}$ are replaced by the reciprocals of polynomials. The integrands
have poles, which can cross the integration path when $k$ is deformed to
complex values. Direct replacements $k\rightarrow k_{\text{M}}$ may jump 
\textit{over} the integration paths, and give incorrect results.

When the theory contains physical particles, besides NL-PVPs, as in the
example (\ref{nlocca}), one proceeds as usual in the physical sector, i.e.,
by Wick rotating the external momenta and keeping the internal ones
Euclidean \cite{efimov}. This procedure is also safe in loops that contain
both physical particles and NL-PVPs.

To conclude this section, the analytic continuation E$\rightarrow $M gives a
well-defined map%
\begin{equation}
\begin{tabular}{ccc}
$\mathcal{M}_{\text{NL}}^{\text{E}}$ & $\overset{\text{E}\rightarrow \text{M}%
}{\longrightarrow }$ & $\mathcal{M}_{\text{NL}}^{\text{M}}$%
\end{tabular}
\label{MapAlEM}
\end{equation}%
between the nonlocal Euclidean theories and their Minkowskian versions. It
is sufficient to replace the external momenta $k$ with their Minkowski
versions $k_{\text{M}}$ inside the loop integrals (or Wick rotate them), and
adjust the overall factor.

\section{Local limit in Euclidean space}

\label{localE}\setcounter{equation}{0}

In this section we study the local limit in Euclidean space, which does not
pose significant challenges.

At the tree level, the assumption (\ref{localass}) ensures that the limit of
the Lagrangian (\ref{nloc}) is the Lagrangian of a local theory:%
\begin{equation}
\mathcal{L}_{\text{NL}}(\lambda )\equiv \frac{\tau }{2}\phi (-p)\lambda
^{-1}Q(\lambda P(p))\phi (p)+\mathcal{O}(\phi ^{3})\underset{\lambda
\rightarrow +\infty }{\rightarrow }\mathcal{L}_{\text{loc}}=\frac{\tau }{2}%
\phi (-p)P(p)\phi (p)+\left. \mathcal{O}(\phi ^{3})\right\vert _{\text{loc}}.
\label{loclim}
\end{equation}%
We assume that the vertices tend to local limits smoothly enough to not
affect our arguments. The theories we have in mind satisfy this assumption,
as we demonstrate in subsection \ref{vertices}.

The propagator tends to the one of the local theory in both cases (\ref{Q}): 
\begin{equation}
\lim_{\lambda \rightarrow +\infty }\frac{\tau \lambda }{Q(\lambda P(p))}=%
\frac{\tau }{P(p)}.  \label{local}
\end{equation}%
It is also straightforward to show that the correlation functions of the
nonlocal theory tend to those of the local theory (\ref{loclim}). Indeed,
the loop integrals just involve the Euclidean domain. There, the values of
the rescaled polynomial $P$ are always real and larger than a positive
number, $\lambda P(0)=\lambda m^{2}$, so the expansion (\ref{expansion}) can
be used to prove the statement.

Take for example (\ref{bub}), with the Green functions (\ref{Gnl}). Its
local limit is%
\begin{equation}
\lim_{\lambda \rightarrow +\infty }\lambda ^{2}\int \frac{\mathrm{d}^{D}p}{%
(2\pi )^{D}}\frac{V_{1}(p,k)V_{2}(p,k)}{Q(\lambda P(p))Q(\lambda P(p-k))}.
\label{locli}
\end{equation}%
Since $k$ is Euclidean, both $P(p)$ and $P(p-k)$ have Euclidean arguments.

Assume that the integral%
\begin{equation}
\mathcal{B}_{\text{L}}(k)\equiv \int \frac{\mathrm{d}^{D}p}{(2\pi )^{D}}%
\frac{V_{1}(p,k)V_{2}(p,k)}{P(p)P(p-k)}  \label{BL}
\end{equation}%
is convergent in the physical dimension $D=4$. The bounds (\ref{bound}) and (%
\ref{bound2}) allow us to apply the dominated convergence theorem and
conclude that the limit (\ref{locli}) gives $\mathcal{B}_{\text{L}}(k)$.

If (\ref{BL}) is not convergent in $D=4$, we use the dimensional
regularization technique. This means that we continue the spacetime
dimension to complex values $D$, move to a domain $\Omega $ where $\mathcal{B%
}_{\text{L}}(k)$ is convergent\footnote{%
If $\mathcal{B}_{\text{L}}(k)$ does not admit a convergence domain $\Omega $%
, it can be split into a finite sum of terms $\mathcal{B}_{\text{L}%
}^{(i)}(k) $ that separately admit convergence domains $\Omega _{i}$. See 
\cite{renormalization}. Then the argument can be applied to each $\mathcal{B}%
_{\text{L}}^{(i)}(k)$ separately.}, take the limit $\lambda \rightarrow
\infty $ there and then analytically continue the result to $D=4$.

This proves that the local limit is well defined in E. In other words, we
have a map%
\begin{equation}
\mathcal{M}_{\text{NL}}^{\text{E}}\overset{\text{loc}}{\longrightarrow }%
\mathcal{M}_{\text{L}}^{\text{E}}  \label{mapE}
\end{equation}%
between nonlocal theories and local theories (\ref{loclim}) in Euclidean
space.

\section{Local limit in Minkowski spacetime}

\label{WickDirect}\setcounter{equation}{0}

The next task is to investigate the local limit of the Minkowskian
correlation functions. In this section we

\begin{enumerate}
\item describe the problems we face;

\item show that, on general grounds, if it exists, the limit cannot contain
ghosts, while it can contain purely virtual particles;

\item study the limit explicitly in correlation functions, concentrating on
the bubble and triangle diagrams.
\end{enumerate}

The outcome is that the local limit is singular with the left option of (\ref%
{Q}), while it gives a local theory with PVPs with the right option (with
some caveats).

\subsection{The problem}

We first note that it is not sufficient to compose the map (\ref{mapE}) with
the continuation E $\rightarrow $ M, because the second step admits a
plurality of outcomes:%
\begin{equation}
\mathcal{M}_{\text{NL}}^{\text{E}}\overset{\text{loc}}{\longrightarrow }%
\mathcal{M}_{\text{L}}^{\text{E}}\overset{\text{E}\rightarrow \text{M}}{%
\longrightarrow }%
\begin{tabular}{ll}
$\nearrow $ & $\mathcal{M}_{\text{L}}^{\text{M-1}}$ \\ 
$\longrightarrow $ & $\mathcal{M}_{\text{L}}^{\text{M-2}}$ \\ 
$\searrow $ & $\mathcal{M}_{\text{L}}^{\text{M-3}}\cdots $%
\end{tabular}
\label{mapEM?}
\end{equation}%
The analytic continuation E $\rightarrow $ M is excluded by unitarity, as we
show below in detail. Among the other options, a special place is reserved
to the \textquotedblleft average continuation\textquotedblright\ \cite{AP},
which defines purely virtual particles. There is a worse possibility,
though: that the local limit in Minkowski spacetime does not even exist. In
particular, we cannot expect that the limit of (\ref{BM})\ is just (\ref{BL}%
) with $k\rightarrow k_{\text{M}}$, which is ill defined.

Ultimately, the task of identifying the right limit $\mathcal{M}_{\text{L}}^{%
\text{M}}$ of $\mathcal{M}_{\text{NL}}^{\text{M}}$ amounts to building the
diagram%
\begin{equation}
\begin{tabular}{ccc}
$\mathcal{M}_{\text{NL}}^{\text{E}}$ & $\overset{\text{loc}_{\text{E}}}{%
\longrightarrow }$ & $\mathcal{M}_{\text{L}}^{\text{E}}$ \\ 
{\footnotesize E} $\downarrow $ {\footnotesize M} &  & {\footnotesize E} $%
\downarrow $ {\footnotesize M} \\ 
$\mathcal{M}_{\text{NL}}^{\text{M}}$ & $\overset{\text{loc}_{\text{M}}}{%
\longrightarrow }$ & $\mathcal{M}_{\text{L}}^{\text{M}}$%
\end{tabular}
\label{mapM}
\end{equation}%
by combining (\ref{MapAlEM}) and (\ref{mapE}) with further maps that close
the bottom right.\ The correct $\mathcal{M}_{\text{L}}^{\text{M}}$ must
follow from the compatibility between the local limit and the continuation E 
$\rightarrow $ M.

\subsection{No-ghost, pro-PVP theorem}

Now we show that $\mathcal{M}_{\text{L}}^{\text{M}}$ cannot contain models
with ghosts, while it can contain theories with purely virtual particles.

Assume for the moment that the vertices $V_{1}$ and $V_{2}$ are identically
one. Making a reflection $p_{4}\rightarrow -p_{4}$ in (\ref{BM}) we infer
that $B_{\text{M}}(k_{\text{M}})=B_{\text{M}}(k_{\text{M}}^{\ast })$.
Because of the property (\ref{reality}) and the overall factor $i$, the
conjugate $B_{\text{M}}^{\ast }(k_{\text{M}})$ coincides with $-B_{\text{M}%
}(k_{\text{M}}^{\ast })$. Hence, $B_{\text{M}}(k_{\text{M}})=-[B_{\text{M}%
}(k_{\text{M}})]^{\ast }$. Given that the real part of $B_{\text{M}}(k_{%
\text{M}})$ vanishes identically, its local limit must vanish as well, if it
exists.

Thus, the local limit of $B_{\text{M}}(k_{\text{M}})$ is purely imaginary.
If it contained propagating particles or ghosts, its real part would be
nontrivial. On the contrary, it is allowed to contain PVPs, because they
give a purely imaginary bubble diagram (check appendix \ref{app1}).

If $V_{1}$ and $V_{2}$ are nontrivial, by unitarity they must be Hermitian
in Minkowski spacetime. The argument just outlined extends straightforwardly
when they are polynomial. More generally, the vertices may involve
incremental ratios of the entire functions $1/Q$, as explained in subsection %
\ref{vertices}. Yet, the operations described above apply to those cases as
well, and lead to the same conclusion.

\bigskip

A more general version of the argument is based on the optical theorem,
which reads $-iT+iT^{\dagger }=TT^{\dagger }$, where $S=1+iT$ is the $S$
matrix and $iT$ collects the loop diagrams in matrix form. Consider the
\textquotedblleft empty\textquotedblright\ theory (\ref{nloc}). We call it
this way, because it does not propagate any degree of freedom, differently
from (\ref{nlocca}). This means $TT^{\dagger }=0$, hence the matrix $T$ is
Hermitian: all the loop diagrams are purely imaginary.

Since Re[$iT$]$=0$ for arbitrary finite $\lambda $, the local limit $\lambda
\rightarrow \infty $ can only return a local theory with Re[$iT$]$=0$, that
is to say, still an empty theory. No physical particles or ghosts can be
generated by the limit. Purely virtual particles are allowed, precisely
because they satisfy Re[$iT$]$=0$.

Coming back to the bubble diagram, we separate the tentative ($\lambda $
independent) local limit from the rest by writing%
\begin{equation}
\hspace{0.01in}\rangle \hspace{-0.18em}{\bigcirc \hspace{-0.18em}}\langle 
\hspace{0.01in}\hspace{0.01in}_{\text{nloc}}=\hspace{0.01in}\rangle \hspace{%
-0.18em}{\bigcirc \hspace{-0.18em}}\langle \hspace{0.01in}\hspace{0.01in}_{%
\text{loc}}+\hspace{0.01in}\rangle \hspace{-0.18em}{\bigcirc \hspace{-0.18em}%
}\langle \hspace{0.01in}\hspace{0.01in}_{\text{rest}}\hspace{0.01in}.
\label{bubb}
\end{equation}%
The rest is supposed to tend to zero when $\lambda $ tends to infinity.

We know that the real part of the left-hand side vanishes, so 
\begin{equation}
0=\text{Re}\left[ \hspace{0.01in}\rangle \hspace{-0.18em}{\bigcirc \hspace{%
-0.18em}}\langle \hspace{0.01in}\hspace{0.01in}_{\text{nloc}}\right] =\text{%
Re}\left[ \hspace{0.01in}\rangle \hspace{-0.18em}{\bigcirc \hspace{-0.18em}}%
\langle \hspace{0.01in}\hspace{0.01in}_{\text{loc}}\right] +\text{Re}\left[ 
\hspace{0.01in}\rangle \hspace{-0.18em}{\bigcirc \hspace{-0.18em}}\langle 
\hspace{0.01in}\hspace{0.01in}_{\text{rest}}\right] .  \label{zero}
\end{equation}%
Assume, ad absurdum, that the local limit adds degrees of freedom. Then the
optical theorem $-iT+iT^{\dagger }=TT^{\dagger }=-2\text{Re}$[$iT$], applied
to the limit itself, tells us that $TT^{\dagger }$ is nonzero. For example,
in the case of the ordinary bubble diagram with circulating physical
particles, formula (\ref{rebub}) gives%
\begin{equation*}
\text{Re}\left[ \hspace{0.01in}\rangle \hspace{-0.18em}{\bigcirc \hspace{%
-0.18em}}\langle \hspace{0.01in}\hspace{0.01in}_{\text{loc}}\right] =-\int 
\frac{\mathrm{d}^{D-1}\mathbf{p}}{(2\pi )^{D-1}}\frac{\pi }{4\omega _{%
\mathbf{p}}\omega _{\mathbf{p}-\mathbf{k}}}\left[ \delta (k^{0}+\omega _{%
\mathbf{p}-\mathbf{k}}+\omega _{\mathbf{p}})+\delta (k^{0}-\omega _{\mathbf{p%
}-\mathbf{k}}-\omega _{\mathbf{p}})\right] .
\end{equation*}

Since the first term on the right hand side of (\ref{zero}) is nonvanishing
and $\lambda $ independent,%
\begin{equation*}
\text{Re}\left[ \hspace{0.01in}\rangle \hspace{-0.18em}{\bigcirc \hspace{%
-0.18em}}\langle \hspace{0.01in}\hspace{0.01in}_{\text{loc}}\right] \neq
0,\qquad \frac{\partial }{\partial \lambda }\text{Re}\left[ \hspace{0.01in}%
\rangle \hspace{-0.18em}{\bigcirc \hspace{-0.18em}}\langle \hspace{0.01in}%
\hspace{0.01in}_{\text{loc}}\right] =0,
\end{equation*}%
the rest is not negligible in the limit $\lambda \rightarrow +\infty $,
which invalidates the assumption.

Assuming, instead, that the local limit gives a theory of PVPs,%
\begin{equation}
\hspace{0.01in}\rangle \hspace{-0.18em}{\bigcirc \hspace{-0.18em}}\langle 
\hspace{0.01in}\hspace{0.01in}_{\text{nloc}}=\hspace{0.01in}\rangle \hspace{%
-0.18em}{\bigcirc \hspace{-0.18em}}\langle \hspace{0.01in}\hspace{0.01in}_{%
\text{PVP}}+\hspace{0.01in}\rangle \hspace{-0.18em}{\bigcirc \hspace{-0.18em}%
}\langle \hspace{0.01in}\hspace{0.01in}_{\text{rest}},  \label{bubbPVP}
\end{equation}%
and recalling that the bubble diagram (\ref{BPVP}) with circulating PVPs is
purely imaginary, the real part of the rest vanishes, 
\begin{equation*}
0=\text{Re}\left[ \hspace{0.01in}\rangle \hspace{-0.18em}{\bigcirc \hspace{%
-0.18em}}\langle \hspace{0.01in}\hspace{0.01in}_{\text{nloc}}\right] =\text{%
Re}\left[ \hspace{0.01in}\rangle \hspace{-0.18em}{\bigcirc \hspace{-0.18em}}%
\langle \hspace{0.01in}\hspace{0.01in}_{\text{rest}}\right] ,
\end{equation*}%
which causes no problem when $\lambda $ tends to $\infty $.

Taking the imaginary part of (\ref{bubbPVP}), we find 
\begin{equation*}
\text{Im}\left[ \hspace{0.01in}\rangle \hspace{-0.18em}{\bigcirc \hspace{%
-0.18em}}\langle \hspace{0.01in}\hspace{0.01in}_{\text{nloc}}\right] =\text{%
Im}\left[ \hspace{0.01in}\rangle \hspace{-0.18em}{\bigcirc \hspace{-0.18em}}%
\langle \hspace{0.01in}\hspace{0.01in}_{\text{PVP}}\right] +\text{Im}\left[ 
\hspace{0.01in}\rangle \hspace{-0.18em}{\bigcirc \hspace{-0.18em}}\langle 
\hspace{0.01in}\hspace{0.01in}_{\text{rest}}\right] .
\end{equation*}%
The left-hand side is nonzero and $\lambda $ dependent. The first term on
the right-hand side is nonzero and $\lambda $ independent. Hence the rest
can be nonzero, $\lambda $ dependent and tend to zero when $\lambda
\rightarrow \infty $ with no contradiction.

We stress once again that these results are predicated on the assumption
that the local limit exists. It turns out that it does exist with the second
choice of (\ref{Q}), but it does not with the first choice. In the rest of
this section we show these facts in detail.

\subsection{Local limit: tree diagrams}

The Euclidean and Minkowskian actions $S_{\text{E}}$ and $S_{\text{M}}$ are
related by $S_{\text{E}}=-iS_{\text{M}}$ (since \textrm{e}$^{-S_{\text{E}%
}}\rightarrow $ \textrm{e}$^{iS_{\text{M}}}$ inside the functional
integral). We know that the integrated coordinates and momenta remain
Euclidean. Thus, the propagator $G_{\text{nl}}^{\text{M}}(p_{\text{M}})$ of
the nonlocal Minkowski theory is (\ref{Gnl}) times $-i$:%
\begin{equation}
G_{\text{nl}}^{\text{M}}(p_{\text{M}})=-\frac{i\tau }{Q(P(p_{\text{M}}))},
\label{GnlM}
\end{equation}%
where $p_{\text{M}}=(-ip^{0},\mathbf{p})$.

At the tree level, formulas (\ref{ess})\ and (\ref{limit})\ give%
\begin{equation}
\lim_{\lambda \rightarrow +\infty }-\frac{i\tau \lambda }{Q(\lambda P(p_{%
\text{M}}))}=\left\{ 
\begin{tabular}{l}
$-\frac{i\tau }{|P(p_{\text{M}})|}$ \\ 
$-\mathcal{P}\frac{i\tau }{P(p_{\text{M}})}$%
\end{tabular}%
\right. 
\begin{tabular}{l}
for $Q(P)=h(P),$ \\ 
for $Q(P)=\frac{h^{2}(P)}{P},$%
\end{tabular}
\label{MinkLim}
\end{equation}%
where $\mathcal{P}$ denotes the Cauchy principal value. The top result
illustrates the problem with the left choice of (\ref{Q}): since we are
assuming that the polynomial $P$ is the one of (\ref{Pp}), the $\lambda
\rightarrow +\infty $ limit does not give an acceptable propagator for a
local theory. The second choice for (\ref{Q}), on the other hand, gives the
answer we expect for a PVP.

A polynomial like (\ref{PHD}) changes the picture. However, we do not have
further results to report on this here.

\subsection{Local limit: bubble diagram}

Now we study the local limit of (\ref{BM}) in detail. We work with the
second option of (\ref{Q}) for the function $Q$ and later comment on the
problems of the first option.

As recalled in appendix \ref{app1}, PVPs are defined by a thoroughlly new
diagrammatics -- i.e., not just the usual diagrams with propagators replaced
by the principal value appearing in (\ref{MinkLim}). Because of this, before
concluding that the local limit with the second choice of (\ref{Q}) gives a
theory with PVPs, we must prove that the loop diagrams somehow
\textquotedblleft know what to do\textquotedblright\ by themselves. That is
to say, for some mysterious reason they are already equipped with the
instructions to implement the diagrammatic rules of \cite%
{AP,diagrammarMio,PVP20}, instead of those that lead to (\ref{wheeler}).
Luckily, this turns out to be true\footnote{%
Besides, we already know that they cannot give (\ref{wheeler}), because
unitarity rules out local limits with ghosts.}.

It is convenient to split the calculation of the bubble diagram in two steps:%
\begin{equation*}
\mathcal{B}_{\text{M}}^{\text{loc}}(k_{\text{M}})=i\lim_{\substack{ \lambda
\rightarrow +\infty  \\ \lambda ^{\prime }\rightarrow +\infty }}\lambda
\lambda ^{\prime }\int \frac{\mathrm{d}^{D}p}{(2\pi )^{D}}\frac{V}{Q(\lambda
^{\prime }P(p))Q(\lambda P(p-k_{\text{M}}))},
\end{equation*}%
where $V$ denotes the vertices. As before, we focus on the propagators and
assume that the vertices behave sufficiently well in the limit, so as not to
invalidate the arguments. We recall that the polynomial $P$ is the one of (%
\ref{Pp}) and $k_{\text{M}}=(-ik^{0},\mathbf{k})$.

Since the loop momentum $p$ is Euclidean, the limit $\lambda ^{\prime
}\rightarrow +\infty $ can be evaluated immediately by means of (\ref{ess}),
(\ref{limit}) or (\ref{local}). We obtain%
\begin{equation}
\mathcal{B}_{\text{M}}^{\text{loc}}(k_{\text{M}})=i\lim_{\lambda \rightarrow
+\infty }\int \frac{\mathrm{d}^{D}p}{(2\pi )^{D}}\frac{V}{p^{2}+m^{2}}\frac{%
\lambda }{Q(\lambda P(p-k_{\text{M}}))}.  \label{BMKM}
\end{equation}

If $k^{0}=0$ the Euclidean and Minkowskian loop integrals coincide, so we
can take the second limit right away, which gives the expected result.

If $k^{0}\neq 0$, we can restrict to the case $k^{0}>0$, by symmetry. Then
the second limit cannot be evaluated directly, since we do not have an easy
control on the behavior of $Q(\lambda P(p-k_{\text{M}}))$ away from the cone 
$\mathcal{C}$.

We briefly outline the strategy of the calculation. The function $Q(\lambda
P(p-k_{\text{M}}))$ is centered in a region translated by $k_{\text{M}}$. We
would like to re-center it on the Euclidean domain by means of an opposite
translation. The translation in question is complex, so it cannot be
expressed as a mere change of variables in the integral.

\begin{figure}[t]
\begin{center}
\includegraphics[width=16truecm]{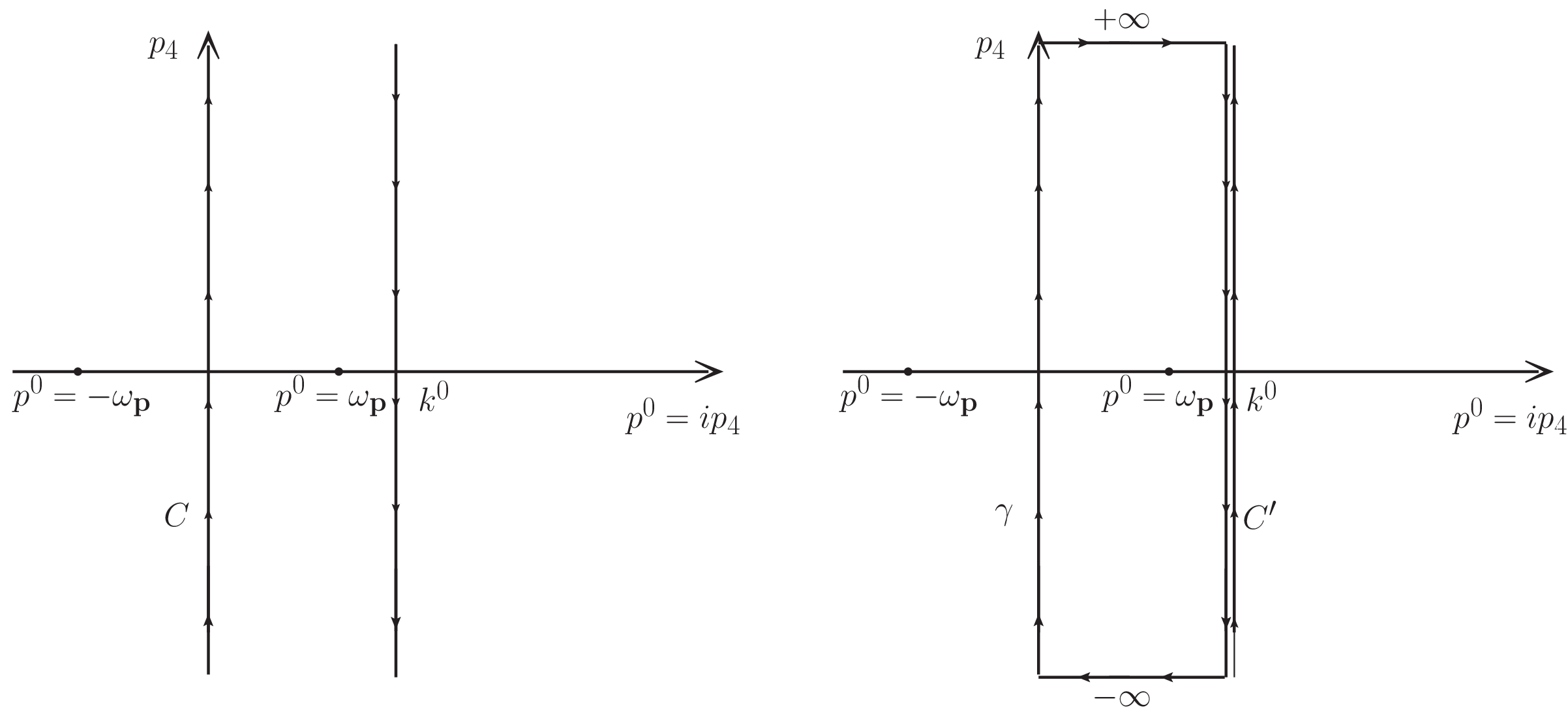}
\end{center}
\par
\vskip -.6truecm
\caption{Dealing with the second limit (assuming $k^{0}>0$)}
\label{Contour}
\end{figure}

We overcome the difficulty by adding and subtracting integration paths.
Consider the $p_{4}$ integral. Its integration domain is the line $C$ (see
fig. \ref{Contour} -- left side). We can rewrite the integral on $C$ as the
sum of the integral on the closed curve $\gamma $, plus the integral on the
line $C^{\prime }$ (fig. \ref{Contour} -- right side). The segments at
infinity do not contribute, since they are located inside the cone $\mathcal{%
C}^{\prime }$ where $\lambda ^{-1}Q(\lambda P(p-k_{\text{M}}))$ converges to 
$P(p-k_{\text{M}})$, as follows from formula (\ref{poly})\footnote{%
Recall that we are using the dimensional regularization. This means that
when an integral is ultraviolet divergent, we split it into a sum of
integrals that admit convergence domains in the complex plane of the
dimension $D$ \cite{renormalization}. Then the arguments are applied to each
of them separately.}.

The integrand of (\ref{BMKM}) has poles at $p^{2}+m^{2}=0$ in the complex $%
p_{4}$ plane. One of them may fall inside the curve $\gamma $. We need to
distinguish the case where this happens from the opposite case. The integral
on $C^{\prime }$, instead, is centered on the Euclidean region, so we can
take the limit $\lambda \rightarrow +\infty $ straightforwardly on it by
means of (\ref{limit}). 

Defining $p^{0}=ip_{4}$, the poles of $p^{2}+m^{2}=0$ are located at $%
p^{0}=\pm \omega _{\mathbf{p}}$, where $\omega _{\mathbf{p}}=\sqrt{\mathbf{p}%
^{2}+m^{2}}$, so $p^{0}=\omega _{\mathbf{p}}$ is inside $\gamma $ if $%
k^{0}>\omega _{\mathbf{p}}$. No pole is inside $\gamma $ if $k^{0}<\omega _{%
\mathbf{p}}$. We can write%
\begin{equation*}
\mathcal{B}_{\text{M}}^{\text{loc}}(k_{\text{M}})=(a)+(b),
\end{equation*}%
where%
\begin{eqnarray}
(a) &=&i\lim_{\lambda \rightarrow +\infty }\int_{\gamma }\frac{\mathrm{d}%
p_{4}}{2\pi }\int_{\omega _{\mathbf{p}}\leqslant k^{0}}\frac{\mathrm{d}^{D-1}%
\mathbf{p}}{(2\pi )^{D-1}}\frac{V}{p^{2}+m^{2}}\frac{\lambda }{Q(\lambda
P(p-k_{\text{M}}))},  \label{a1} \\
(b) &=&i\lim_{\lambda \rightarrow +\infty }\int_{C^{\prime }}\frac{\mathrm{d}%
p_{4}}{2\pi }\int \frac{\mathrm{d}^{D-1}\mathbf{p}}{(2\pi )^{D-1}}\frac{V}{%
p^{2}+m^{2}}\frac{\lambda }{Q(\lambda P(p-k_{\text{M}}))}.  \label{b1}
\end{eqnarray}

Let us begin by evaluating $(a)$ and $(b)$ in the simplified case $\mathbf{k}%
=m=0$. The residue theorem gives\footnote{%
Note that the integral is in $p_{4}=-ip^{0}$, not $p^{0}$, so there is an
extra factor $-i$.}%
\begin{equation}
(a)=i\lim_{\lambda \rightarrow +\infty }\int_{|\mathbf{p}|\leqslant k^{0}}%
\frac{\mathrm{d}^{D-1}\mathbf{p}}{(2\pi )^{D-1}}\frac{V}{2|\mathbf{p}|}\frac{%
\lambda }{Q(\lambda k^{0}(2|\mathbf{p}|-k^{0}))}.  \label{(a)}
\end{equation}%
Since the argument of $Q$ is real, we can take the limit $\lambda
\rightarrow +\infty $ by means of formula (\ref{limit}). We obtain%
\begin{equation}
(a)=-i\mathcal{P}\int_{|\mathbf{p}|\leqslant k^{0}}\frac{\mathrm{d}^{D-1}%
\mathbf{p}}{(2\pi )^{D-1}}\frac{V}{2|\mathbf{p}|}\frac{1}{k^{0}(k^{0}-2|%
\mathbf{p}|)}.  \label{aa}
\end{equation}

The integral $(b)$ is centered on the Euclidean domain. It is convenient to
make this fact apparent by relabeling the integration variable $p_{4}$
(which is not a change of variables) and writing%
\begin{equation}
(b)=i\lim_{\lambda \rightarrow +\infty }\int \frac{\mathrm{d}^{D}p}{(2\pi
)^{D}}\frac{V}{(p+k_{\text{M}})^{2}}\frac{\lambda }{Q(\lambda P(p))}.
\label{bb}
\end{equation}%
The limit $\lambda \rightarrow +\infty $ is now straightforward and gives%
\begin{equation}
(b)=i\int \frac{\mathrm{d}^{D}p}{(2\pi )^{D}}\frac{V}{(p+k_{\text{M}})^{2}}%
\frac{1}{p^{2}}.  \label{fbubble}
\end{equation}%
Note that two more poles are created by the limit. Moreover,(\ref{fbubble})
does not coincide with the loop integral of the usual bubble diagram,
because when we apply the residue theorem to the $p_{4}$ integral, we end up
including different poles.

Closing the path on the right side, we encircle two poles ($p^{0}=|\mathbf{p}%
|-k^{0}$ and $p^{0}=|\mathbf{p}|$) for $k^{0}<|\mathbf{p}|$ and one ($p^{0}=|%
\mathbf{p}|$) for $k^{0}>|\mathbf{p}|$. The result is%
\begin{equation}
(b)=-i\int_{k^{0}<|\mathbf{p}|}\frac{\mathrm{d}^{D-1}\mathbf{p}}{(2\pi
)^{D-1}}\frac{V}{2|\mathbf{p}|}\frac{1}{k^{0}(k^{0}-2|\mathbf{p}|)}-i\int 
\frac{\mathrm{d}^{D-1}\mathbf{p}}{(2\pi )^{D-1}}\frac{V}{2|\mathbf{p}|}\frac{%
1}{k^{0}(k^{0}+2|\mathbf{p}|)}.  \label{posi}
\end{equation}%
This expression does not need particular prescriptions, since the integrand
is never singular.

Summing to $(a)$, we find the total%
\begin{eqnarray}
\mathcal{B}_{\text{M}}^{\text{loc}}(k_{\text{M}}) &=&-i\mathcal{P}\int \frac{%
\mathrm{d}^{D-1}\mathbf{p}}{(2\pi )^{D-1}}\frac{V}{2|\mathbf{p}|}\frac{1}{%
k^{0}(k^{0}-2|\mathbf{p}|)}-i\int \frac{\mathrm{d}^{D-1}\mathbf{p}}{(2\pi
)^{D-1}}\frac{V}{2|\mathbf{p}|}\frac{1}{k^{0}(k^{0}+2|\mathbf{p}|)}  \notag
\\
&=&i\mathcal{P}\int \frac{\mathrm{d}^{D-1}\mathbf{p}}{(2\pi )^{D-1}}\frac{V}{%
4|\mathbf{p}|^{2}}\left( \frac{1}{k^{0}+2|\mathbf{p}|}-\frac{1}{k^{0}-2|%
\mathbf{p}|}\right) ,  \label{reso}
\end{eqnarray}%
which is precisely the result obtained in the case of PVPs (apart from the
factor $V$), as given in formula (\ref{BPVP}).

We have obtained the result (\ref{reso}) by adopting the second option of (%
\ref{Q}). Let us see what happens, instead, when we adopt the first option.
Everything proceeds as above in $(b)$, because the arguments of $Q$\ are
positive. The difference is visible in $(a)$, where we obtain%
\begin{equation}
(a)=i\int_{|\mathbf{p}|\leqslant k^{0}}\frac{\mathrm{d}^{D-1}\mathbf{p}}{%
(2\pi )^{D-1}}\frac{V}{2|\mathbf{p}|}\frac{1}{k^{0}|k^{0}-2|\mathbf{p}||},
\label{singula}
\end{equation}%
instead of (\ref{aa}), which is clearly singular. Thus, the local limit is
not well defined with the first option of (\ref{Q}) in Minkowski spacetime.

For completeness, let us treat the general case $\mathbf{k}\neq 0$, $m\neq 0$
with the second option (\ref{Q}) for $Q$. The residue theorem gives (\ref{a1}%
) again, but now on the pole $p^{0}=\omega _{\mathbf{p}}$ we have%
\begin{equation}
(p-k_{\text{M}})^{2}+m^{2}=-(k^{0}-\omega _{\mathbf{p}}-\omega _{\mathbf{p}-%
\mathbf{k}})(k^{0}-\omega _{\mathbf{p}}+\omega _{\mathbf{p}-\mathbf{k}}).
\label{pmk}
\end{equation}%
What is important is that the argument of $Q$ remains real, so we can still
use formula (\ref{limit}) to evaluate the limit $\lambda \rightarrow +\infty 
$. The result is%
\begin{equation}
(a)=-i\mathcal{P}\int \frac{\mathrm{d}^{D-1}\mathbf{p}}{(2\pi )^{D-1}}\frac{%
V\theta (k^{0}-\omega _{\mathbf{p}})}{4\omega _{\mathbf{p}}\omega _{\mathbf{p%
}-\mathbf{k}}}\left( \frac{1}{k^{0}-\omega _{\mathbf{p}}-\omega _{\mathbf{p}-%
\mathbf{k}}}-\frac{1}{k^{0}-\omega _{\mathbf{p}}+\omega _{\mathbf{p}-\mathbf{%
k}}}\right) .  \label{aaa}
\end{equation}%
The singularity due to the first term inside the parentheses is regulated by
the principal value inherited from (\ref{limit}). Instead, the second term
is nonsingular for $\omega _{\mathbf{p}}\leqslant k^{0}$.

Centering the $(b)$ integral as in (\ref{bb}), taking $\lambda $ to infinity
and translating $\mathbf{p}$ back to $\mathbf{p}-\mathbf{k}$, we obtain%
\begin{equation}
(b)=i\int \frac{\mathrm{d}^{D}p}{(2\pi )^{D}}\frac{V}{(p^{0}+k^{0}-\omega _{%
\mathbf{p}})(p^{0}+k^{0}+\omega _{\mathbf{p}})}\frac{1}{(p^{0}-\omega _{%
\mathbf{p}-\mathbf{k}})(p^{0}+\omega _{\mathbf{p}-\mathbf{k}})}.
\label{fbubble2}
\end{equation}%
Repeating the argument above, we close the path on the right side, thereby
picking two poles ($p^{0}=\omega _{\mathbf{p}}-k^{0}$ and $p^{0}=\omega _{%
\mathbf{p}-\mathbf{k}}$) for $k^{0}<\omega _{\mathbf{p}}$ and one ($%
p^{0}=\omega _{\mathbf{p}-\mathbf{k}}$) for $k^{0}>\omega _{\mathbf{p}}$.
The result is%
\begin{equation*}
(b)=-i\int \frac{\mathrm{d}^{D-1}\mathbf{p}}{(2\pi )^{D-1}}\frac{V}{4\omega
_{\mathbf{p}}\omega _{\mathbf{p}-\mathbf{k}}}\left( \frac{\theta (\omega _{%
\mathbf{p}}-k^{0})}{k^{0}-\omega _{\mathbf{p}}-\omega _{\mathbf{p}-\mathbf{k}%
}}+\frac{\theta (k^{0}-\omega _{\mathbf{p}})}{k^{0}-\omega _{\mathbf{p}%
}+\omega _{\mathbf{p}-\mathbf{k}}}-\frac{1}{k^{0}+\omega _{\mathbf{p}%
}+\omega _{\mathbf{p}-\mathbf{k}}}\right) .
\end{equation*}%
Again, the integrand is never singular.

Summing the outcome to the $(a)$ of (\ref{aaa}), we finally get%
\begin{equation*}
\mathcal{B}_{\text{M}}^{\text{loc}}(k_{\text{M}})=-i\mathcal{P}\int \frac{%
\mathrm{d}^{D-1}\mathbf{p}}{(2\pi )^{D-1}}\frac{V}{4\omega _{\mathbf{p}%
}\omega _{\mathbf{p}-\mathbf{k}}}\left( \frac{1}{k^{0}-\omega _{\mathbf{p}%
}-\omega _{\mathbf{p}-\mathbf{k}}}-\frac{1}{k^{0}+\omega _{\mathbf{p}%
}+\omega _{\mathbf{p}-\mathbf{k}}}\right) ,
\end{equation*}%
which is the same as in the case of PVPs, formula (\ref{BPVP}).

If one particle circulating in the loop is physical, like $\varphi $ in (\ref%
{nlocca}), and the other one is $\phi $ with the second option of (\ref{Q})
for $Q$, the result does not change. The simplest parametrization of the
circulating momenta gives (\ref{BMKM}) directly, whence the rest follows as
before. If we start from the parametrization 
\begin{equation}
\mathcal{B}_{\text{M}}^{\text{loc}}(k_{\text{M}})=i\lim_{\lambda \rightarrow
+\infty }\int \frac{\mathrm{d}^{D}p}{(2\pi )^{D}}\frac{V}{(p+k_{\text{M}%
})^{2}+m^{2}}\frac{\lambda }{Q(\lambda P(p))},  \label{bu}
\end{equation}%
we have to note that the integration on $p_{4}$ must include complex values
in order to leave the right pole on one side and the left pole on the other
side. Then the argument of $Q$ is not everywhere real, so the limit $\lambda
\rightarrow +\infty $ cannot be taken directly. On the other hand, one can
easily see that (\ref{bu}) is equivalent to (\ref{BMKM}), because the
difference is a closed path that encircles no pole.

\subsection{Local limit:\ triangle diagram}

Now we study the loop integral%
\begin{equation}
\mathcal{T}_{\text{M}}(k_{\text{M}},q_{\text{M}})=i\int \frac{\mathrm{d}^{D}p%
}{(2\pi )^{D}}V(p,k_{\text{M}},q_{\text{M}})G_{\text{nl}}(p)G_{\text{nl}%
}(p-k_{\text{M}})G_{\text{nl}}(p-q_{\text{M}})  \label{TT}
\end{equation}%
of the triangle diagram with the second option of (\ref{Q}) for $Q$. After
possibly a translation of the internal momentum $p$, and assuming that the
external momenta are generic, we can arrange the expression so that, say, $%
q^{0}>k^{0}>0$.

We introduce the parameter $\lambda $ and split the local limit $\lambda
\rightarrow +\infty $ in three steps. First, we take the limit inside $G_{%
\text{nl}}(p)$, then in $G_{\text{nl}}(p-k_{\text{M}})$ and finally in $G_{%
\text{nl}}(p-q_{\text{M}})$. The calculation will tell us to what extent it
is legitimate to do so.

As before, the first limit is straightforward, since $G_{\text{nl}}(p)$ does
not depend on $k_{\text{M}}$ and $q_{\text{M}}$. We remain with%
\begin{equation*}
\mathcal{T}_{\text{M}}^{\text{loc}}(k_{\text{M}},q_{\text{M}%
})=i\lim_{\lambda ^{\prime }\rightarrow +\infty }\lim_{\lambda \rightarrow
+\infty }\int \frac{\mathrm{d}^{D}p}{(2\pi )^{D}}\frac{V}{p^{2}+m^{2}}\frac{%
\lambda ^{\prime }}{Q(\lambda ^{\prime }P(p-k_{\text{M}}))}\frac{\lambda }{%
Q(\lambda P(p-q_{\text{M}}))}.
\end{equation*}

Now we use the residue theorem to integrate the loop energy $p^{4}=-ip^{0}$
along the closed curve $\gamma $ made of the lines $p^{0}=0$ and $p^{0}=k^{0}
$, plus segments at infinity. On the pole $p^{0}=\omega _{\mathbf{p}}$,
which contributes for $\omega _{\mathbf{p}}<k^{0}$, the arguments of the
functions $Q$ are real, so we can use formula (\ref{limit}). We then obtain
the first contribution, which is%
\begin{equation*}
(a)=-i\int \frac{\mathrm{d}^{D-1}\mathbf{p}}{(2\pi )^{D-1}}\frac{\theta
(k^{0}-\omega _{\mathbf{p}})}{8\omega _{\mathbf{p}}\omega _{\mathbf{p-k}%
}\omega _{\mathbf{p-q}}}\mathcal{Q}^{12}\mathcal{Q}^{13},
\end{equation*}%
where%
\begin{equation*}
\mathcal{Q}^{ab}=\mathcal{P}^{ab}-\mathcal{P}\frac{1}{e_{a}-e_{b}-\omega
_{a}+\omega _{b}},\qquad \mathcal{P}^{ab}=\mathcal{P}\frac{1}{%
e_{a}-e_{b}-\omega _{a}-\omega _{b}},
\end{equation*}%
and the subscripts $a,b,\ldots $ range over the values $1$, $2$ and $3$,
while $\{e_{a}\}=\{0,-k^{0},-q^{0}\}$, $\{\omega _{a}\}=\{\omega _{\mathbf{p}%
},\omega _{\mathbf{p-k}},\omega _{\mathbf{p-q}}\}$.

We are left with the integral on the line $p^{0}=k^{0}$, which we center on $%
p^{0}=0$ by means of a relabelling of the loop energy. After that, the limit 
$\lambda ^{\prime }\rightarrow +\infty $ acts on $\lambda ^{\prime
}/Q(\lambda ^{\prime }P(p))$ and becomes straightforward. We remain with 
\begin{equation}
(b)=i\lim_{\lambda \rightarrow +\infty }\int \frac{\mathrm{d}^{D}p}{(2\pi
)^{D}}\frac{V}{(p^{2}+m^{2})((p+k_{\text{M}})^{2}+m^{2})}\frac{\lambda }{%
Q(\lambda P(p+k_{\text{M}}-q_{\text{M}}))}.  \label{bi}
\end{equation}

At this point, we write $(b)=(b_{1})+(b_{2})$, where $(b_{1})$ is the
integral on the closed curve $\gamma ^{\prime }$ made by the lines $p^{0}=0$
and $p^{0}=q^{0}-k^{0}$, plus segments at infinity, and $(b_{2})$ is the
integral on the line $p^{0}=q^{0}-k^{0}$.

Using the residue theorem once more on $(b_{1})$, we obtain an integral on $%
\mathbf{p}$ that we do not report here. We just remark that, before taking
the limit $\lambda \rightarrow \infty $, its integrand is regular.
Specifically, the residues at the poles mutually compensate for every $%
\lambda <\infty $. This means that we can adopt the prescription we want for
those poles. We choose the principal value. Then, we take the limit $\lambda
\rightarrow \infty $ by means of (\ref{limit}), noting that the arguments of 
$Q$ are real on the poles. We find that the contribution of $\gamma ^{\prime
}$ is%
\begin{equation*}
(b_{1})=-i\int \frac{\mathrm{d}^{D-1}\mathbf{p}}{(2\pi )^{D-1}}\frac{\theta
(q^{0}-\omega _{\mathbf{p}})\theta (\omega _{\mathbf{p}}-k^{0})\mathcal{Q}%
^{12}\mathcal{Q}^{13}+\theta (q^{0}-k^{0}-\omega _{\mathbf{p-k}})\mathcal{Q}%
^{23}\mathcal{Q}^{21}}{8\omega _{\mathbf{p}}\omega _{\mathbf{p-k}}\omega _{%
\mathbf{p-q}}}.
\end{equation*}

Centering the integral $(b_{2})$ on $p^{0}=0$ by means of a further
relabelling of the loop energy, we obtain%
\begin{equation}
(b_{2})=i\lim_{\lambda \rightarrow +\infty }\int \frac{\mathrm{d}^{D}p}{%
(2\pi )^{D}}\frac{V}{((p+q_{\text{M}})^{2}+m^{2})((p-k_{\text{M}}+q_{\text{M}%
})^{2}+m^{2})}\frac{\lambda }{Q(\lambda P(p))}.  \label{bb2}
\end{equation}%
Now the limit $\lambda \rightarrow +\infty $ acts on $\lambda /Q(\lambda
P(p))$. Nevertheless, we cannot evaluate it directly by means of (\ref{limit}%
), because if we do so, we find an unprescribed expression:%
\begin{equation}
(b_{2})=i\int \frac{\mathrm{d}^{D}p}{(2\pi )^{D}}\frac{V}{%
(p^{2}+m^{2})((p-k_{\text{M}}+q_{\text{M}})^{2}+m^{2})((p+q_{\text{M}%
})^{2}+m^{2})}.  \label{jump}
\end{equation}%
Recall that $p$ is Euclidean, so this formula does not correspond (for,
e.g., $V=1$) to the triangle diagram of local theories, due to the different
sets of contributing poles. Using the residue theorem, the result of (\ref%
{jump}) is%
\begin{equation}
(b_{2})=-i\int \frac{\mathrm{d}^{D-1}\mathbf{p}}{(2\pi )^{D-1}}\frac{\theta
(\omega _{\mathbf{p}}-q^{0})\mathcal{\hat{Q}}^{12}\mathcal{\hat{Q}}%
^{13}+\theta (k^{0}-q^{0}+\omega _{\mathbf{p-k}})\mathcal{\hat{Q}}^{23}%
\mathcal{\hat{Q}}^{21}+\mathcal{\hat{Q}}^{32}\mathcal{\hat{Q}}^{31}}{8\omega
_{\mathbf{p}}\omega _{\mathbf{p-k}}\omega _{\mathbf{p-q}}},  \label{b2}
\end{equation}%
where $\mathcal{\hat{Q}}$ is the \textquotedblleft unprescribed\ $\mathcal{Q}
$\textquotedblright , that is to say, the same as $\mathcal{Q}$ without the
principal-value prescription.

Among the other things, the integrand of (\ref{b2}) involves an expression
like%
\begin{equation}
\frac{1}{xy}-\frac{1}{x(x+y)}-\frac{1}{y(x+y)},  \label{sumxy}
\end{equation}%
with $x=\omega _{\mathbf{p}}-\omega _{\mathbf{p}-\mathbf{q}}-q^{0}$ and $%
y=k^{0}-\omega _{\mathbf{p}}+\omega _{\mathbf{p}-\mathbf{k}}$. We focus on
the region around $x=y=0$, where the $\theta $ functions appearing in some
numerators leading to (\ref{sumxy}) are equal to one. Since the sum (\ref%
{sumxy}) is unprescribed, we do not know whether it is zero or not. For
example, it is zero if we slightly move $x$ and $y$ to complex values, but
it is nonzero if we take the principal value, due to the identity \cite%
{diagrammarMio} 
\begin{equation}
\mathcal{P}\left( \frac{1}{xy}-\frac{1}{x(x+y)}-\frac{1}{y(x+y)}\right)
=-\pi ^{2}\delta (x)\delta (y).  \label{Pide}
\end{equation}%
Note that the double delta function on the right-hand side does not appear
in Feynman diagrams (check \cite{diagrammarMio} for details).

The problem just outlined indicates that we have jumped to (\ref{jump}) too
quickly, since the integral of (\ref{bb2}) is well defined before taking the
limit. To avoid the trouble, we can evaluate (\ref{bb2}) as follows. First,
we move the external energies slightly away from the real axis: $%
k^{0}\rightarrow k^{0}+i\epsilon $, $q^{0}\rightarrow q^{0}+i\epsilon
^{\prime }$, with $\epsilon $ and $\epsilon ^{\prime }$ real (not
necessarily positive) and small. Then we use (\ref{limit}) to evaluate the
limit $\lambda \rightarrow +\infty $. When we apply the residue theorem, we
find expressions like (\ref{sumxy}) with $x\rightarrow x-i\epsilon ^{\prime
} $ and $y\rightarrow y+i\epsilon $, which do vanish.

Using the principal value and subtracting the right-hand side of (\ref{Pide}%
), we can write 
\begin{eqnarray*}
(b_{2}) &=&-i\int \frac{\mathrm{d}^{D-1}\mathbf{p}}{(2\pi )^{D-1}}V\frac{%
\theta (\omega _{\mathbf{p}}-q^{0})\mathcal{Q}^{12}\mathcal{Q}^{13}+\theta
(k^{0}-q^{0}+\omega _{\mathbf{p-k}})\mathcal{Q}^{23}\mathcal{Q}^{21}+%
\mathcal{Q}^{32}\mathcal{Q}^{31}}{8\omega _{\mathbf{p}}\omega _{\mathbf{p-k}%
}\omega _{\mathbf{p-q}}} \\
&&+i\pi ^{2}\int \frac{\mathrm{d}^{D-1}\mathbf{p}}{(2\pi )^{D-1}}V\frac{%
\delta (q^{0}-\omega _{\mathbf{p}}+\omega _{\mathbf{p-q}})\delta
(k^{0}-\omega _{\mathbf{p}}+\omega _{\mathbf{p-k}})}{8\omega _{\mathbf{p}%
}\omega _{\mathbf{p-k}}\omega _{\mathbf{p-q}}}.
\end{eqnarray*}

The total gives%
\begin{eqnarray*}
\mathcal{T}_{\text{M}}^{\text{loc}}(k_{\text{M}},q_{\text{M}})
&=&(a)+(b)=-i\int \frac{\mathrm{d}^{D-1}\mathbf{p}}{(2\pi )^{D-1}}V\frac{%
\mathcal{Q}^{12}\mathcal{Q}^{13}+\mathcal{Q}^{23}\mathcal{Q}^{21}+\mathcal{Q}%
^{32}\mathcal{Q}^{31}}{8\omega _{\mathbf{p}}\omega _{\mathbf{p-k}}\omega _{%
\mathbf{p-q}}} \\
&&+i\pi ^{2}\int \frac{\mathrm{d}^{D-1}\mathbf{p}}{(2\pi )^{D-1}}V\frac{%
\delta (q^{0}-\omega _{\mathbf{p}}+\omega _{\mathbf{p-q}})\delta
(k^{0}-\omega _{\mathbf{p}}+\omega _{\mathbf{p-k}})}{8\omega _{\mathbf{p}%
}\omega _{\mathbf{p-k}}\omega _{\mathbf{p-q}}}.
\end{eqnarray*}%
Rearranging it by means of (\ref{Pide}), we obtain%
\begin{eqnarray}
\mathcal{T}_{\text{M}}^{\text{loc}}(k_{\text{M}},q_{\text{M}}) &=&-i\int 
\frac{\mathrm{d}^{D-1}\mathbf{p}}{(2\pi )^{D-1}}V\frac{\mathcal{P}^{12}%
\mathcal{P}^{13}+\mathcal{P}^{21}\mathcal{P}^{31}+\text{cycl}}{8\omega _{%
\mathbf{p}}\omega _{\mathbf{p-k}}\omega _{\mathbf{p-q}}}  \notag \\
&&+i\pi ^{2}\int \frac{\mathrm{d}^{D-1}\mathbf{p}}{(2\pi )^{D-1}}V\frac{%
\delta (q^{0}-\omega _{\mathbf{p}}-\omega _{\mathbf{p-q}})\delta
(q^{0}-k^{0}-\omega _{\mathbf{p-k}}-\omega _{\mathbf{p-q}})}{8\omega _{%
\mathbf{p}}\omega _{\mathbf{p-k}}\omega _{\mathbf{p-q}}}.\quad  \label{result}
\end{eqnarray}

The expected result for a triangle with circulating PVPs is just the first
line \cite{diagrammarMio}. The second line is not correct.

The original integral (\ref{TT}) is symmetric (up to $V$) under exchanges of
the three energies and the reflection $e\rightarrow -e$, but the second line
of (\ref{result}) is not. Indeed, the symmetric expression%
\begin{equation*}
\sum_{a\neq b\neq c\neq a}\delta (e_{a}-e_{b}-\omega _{a}-\omega _{b})\delta
(e_{a}-e_{c}-\omega _{a}-\omega _{c})+(e\rightarrow -e),
\end{equation*}%
specialized to our case, gives%
\begin{equation}
\delta (q^{0}-\omega _{\mathbf{p}}-\omega _{\mathbf{p-q}})\delta
(q^{0}-k^{0}-\omega _{\mathbf{p-k}}-\omega _{\mathbf{p-q}})+\delta
(k^{0}-\omega _{\mathbf{p}}-\omega _{\mathbf{p-k}})\delta (q^{0}-\omega _{%
\mathbf{p}}-\omega _{\mathbf{p-q}}),  \label{doubledeltas}
\end{equation}%
but the second term is missing in (\ref{result}). This means that the
calculation fails in the multi-threshold sector, which is the one made by
the double deltas. Splitting the limit $\lambda \rightarrow \infty $ into
three distinct limits is only accurate up to those terms.

The first line of (\ref{result}) is enough to show that there are no
propagating degrees of freedom. Indeed, the outcome is purely imaginary. In
particular, the single-delta contributions (those which contribute to the
optical theorem and highlight the degrees of freedom propagating on-shell
inside the diagram) are completely missing.

One may object that corrections proportional to (\ref{doubledeltas}) are
also not acceptable, because they are on-shell. Currently, we lack
computational tools that are powerful enough to determine whether they are
actually present or not. At any rate, we can explain how we should proceed
if they were. Basically, we would be forced to advocate the inherent
arbitrariness of the nonlocal deformation to compensate for the extra
contributions. The Lagrangian that gives the correct local limit, including
the multi-threshold interaction sector, would have to be adjusted along the
way by including suitable nonlocal, finite counter-vertices.

\section{Asymptotically local quantum gravity}

\label{alqg}\setcounter{equation}{0}

In this section we explain how to deform the (local) theory of quantum
gravity with purely virtual particles (PVP-QG) \cite{LWgrav} into a unitary,
nonlocal theory that tends to it in the local limit. We call the latter
\textquotedblleft asymptotically local quantum gravity\textquotedblright\
(AL-QG). We work in Minkowski spacetime.

\subsection{PVP-QG}

The PVP-QG theory coupled to matter is described by the higher-derivative
action \cite{Absograv} 
\begin{equation}
S_{\text{QG}}(g,\Phi )=-\frac{1}{16\pi G}\int \mathrm{d}^{4}x\sqrt{-g}\left(
2\Lambda +R-\frac{R^{2}}{6m_{\phi }^{2}}+\frac{\eta }{2m_{\chi }^{2}}C_{\mu
\nu \rho \sigma }C^{\mu \nu \rho \sigma }\right) +S_{m}(g,\Phi ),
\label{SQG}
\end{equation}%
where $m_{\phi }$ is the mass of the inflaton $\phi $ (introduced below), $%
m_{\chi }$ is the mass of the spin-2 massive mode $\chi _{\mu \nu }$ (due to
the square of the Weyl tensor $C_{\mu \nu \rho \sigma }$), which must be
treated as a PVP, $\Phi $ denotes the matter fields, with action $%
S_{m}(g,\Phi )$, and $\eta =m_{\chi }^{2}(3m_{\phi }^{2}+4\Lambda )/(m_{\phi
}^{2}(3m_{\chi }^{2}-2\Lambda ))$ is a parameter very close to one. The
simplest option is to take $S_{m}$ equal to the covariantized action of the
standard model, equipped with the nonminimal couplings allowed by power
counting. The resulting theory is renormalizable\footnote{%
Renormalization works exactly as in the Stelle theory \cite{Stelle}, where
the spin-2 massive field is quantized in a conventional way and propagates a
ghost.} and unitary.

The crucial point is the treatment of $\chi _{\mu \nu }$ as a PVP. To
clarify how this works it is convenient to introduce $\phi $ and $\chi _{\mu
\nu }$ explicitly as extra fields by eliminating the higher derivatives. The
result is \cite{Absograv}%
\begin{equation}
S_{\text{QG}}(\tilde{g},\phi ,\chi ,\Phi )=\tilde{S}_{\text{HE}}(\tilde{g}%
)+S_{\phi }(\tilde{g}+\psi ,\phi )+S_{\chi }(\tilde{g},\chi )+S_{m}(\tilde{g}%
\mathrm{e}^{\kappa _{\phi }\phi }+\psi \mathrm{e}^{\kappa _{\phi }\phi
},\Phi ).  \label{sew}
\end{equation}%
The relation between the old metric $g_{\mu \nu }$ and the new metric $%
\tilde{g}_{\mu \nu }$ reads%
\begin{equation}
g_{\mu \nu }=(\tilde{g}_{\mu \nu }+\psi _{\mu \nu })\mathrm{e}^{\kappa
_{\phi }\phi },\qquad \psi _{\mu \nu }\equiv 2\kappa _{\chi }\chi _{\mu \nu
}+\kappa _{\chi }^{2}\left( \chi _{\mu \nu }\chi _{\rho \sigma }\tilde{g}%
^{\rho \sigma }-2\chi _{\mu \rho }\chi _{\nu \sigma }\tilde{g}^{\rho \sigma
}\right) ,  \label{redef}
\end{equation}%
and the constants are%
\begin{equation*}
\kappa _{\phi }=\frac{m_{\phi }\sqrt{16\pi G}}{\sqrt{4\Lambda +3m_{\phi }^{2}%
}},\qquad \kappa _{\chi }=\sqrt{8\pi \tilde{G}},\qquad \tilde{G}=\frac{G}{%
\eta }.
\end{equation*}%
Moreover,%
\begin{equation*}
\tilde{S}_{\text{HE}}(g)=-\frac{1}{16\pi \tilde{G}}\int \mathrm{d}^{4}x\sqrt{%
-g}\left( 2\Lambda +R\right) ,
\end{equation*}%
is the Einstein-Hilbert action with the redefined Newton constant, 
\begin{equation}
S_{\phi }(g,\phi )=\frac{1}{2}\int \mathrm{d}^{4}x\sqrt{-g}\left[ D_{\mu
}\phi D^{\mu }\phi -\frac{m_{\phi }^{2}}{\kappa _{\phi }^{2}}\left( 1-%
\mathrm{e}^{\kappa _{\phi }\phi }\right) ^{2}\right] ,  \label{sphi}
\end{equation}%
is the inflaton action, and%
\begin{equation}
S_{\chi }(\tilde{g},\chi )=\tilde{S}_{\text{HE}}(\tilde{g}+\psi )-\tilde{S}_{%
\text{HE}}(\tilde{g})+\int \mathrm{d}^{4}x\left[ \frac{m_{\chi }^{2}}{2}%
\sqrt{-g}(\chi _{\mu \nu }\chi ^{\mu \nu }-\chi ^{2})-2\kappa _{\chi }%
\hspace{0.01in}\chi _{\mu \nu }\frac{\delta \tilde{S}_{\text{HE}}(g)}{\delta
g_{\mu \nu }}\right] _{g\rightarrow \tilde{g}+\psi }  \label{spsi}
\end{equation}%
is the $\chi _{\mu \nu }$ action. Specifically, one finds%
\begin{equation}
S_{\chi }(g,\chi )=-S_{\text{PF}}(g,\chi ,m_{\chi }^{2})+S_{\chi
}^{(>2)}(g,\chi ),  \label{scc}
\end{equation}%
where%
\begin{eqnarray}
S_{\text{PF}}(g,\chi ,m_{\chi }^{2}) &=&\frac{1}{2}\int \mathrm{d}^{4}x\sqrt{%
-g}\left[ D_{\rho }\chi _{\mu \nu }D^{\rho }\chi ^{\mu \nu }-D_{\rho }\chi
D^{\rho }\chi +2D_{\mu }\chi ^{\mu \nu }D_{\nu }\chi -2D_{\mu }\chi ^{\rho
\nu }D_{\rho }\chi _{\nu }^{\mu }\right.   \notag \\
&&\left. -m_{\chi }^{2}(\chi _{\mu \nu }\chi ^{\mu \nu }-\chi ^{2})+R^{\mu
\nu }(\chi \chi _{\mu \nu }-2\chi _{\mu \rho }\chi _{\nu }^{\rho })\right] 
\label{SPF}
\end{eqnarray}%
is the covariantized Pauli-Fierz action with a nonminimal term, and $S_{\chi
}^{(>2)}(g,\chi )$ are corrections at least cubic in $\chi $.

The theory (\ref{sew}) is renormalizable, but not manifestly. This means
that, when we calculate its Feynman diagrams, \textquotedblleft
miraculous\textquotedblright\ cancellations make it possible to subtract the
divergences by means of field redefinitions and renormalizations of the
parameters already contained in (\ref{sew}).

The crucial problem is the minus sign in front of $S_{\text{PF}}$. If $\chi
_{\mu \nu }$ is treated conventionally, it propagates a ghost, and unitarity
is violated. For analogous reasons, the theory (\ref{SQG})-(\ref{sew}) is
not acceptable as a classical theory.

The situation changes radically when $\chi _{\mu \nu }$ is understood as a
PVP. The field $\chi _{\mu \nu }$ is projected away by integrating it out
according to the diagrammatic rules of PVPs, briefly recalled in the
appendix. No $\chi _{\mu \nu }$ external legs are considered, and the
modified diagrams guarantee that $\chi _{\mu \nu }$ does not give on-shell
contributions to the radiative corrections.

Expanding around the flat-space metric $\eta _{\mu \nu }=$diag$(1,-1,-1,-1)$%
, the $\chi _{\mu \nu }$ propagator of (\ref{sew}) is%
\begin{equation}
\langle \chi _{\mu \nu }(p)\hspace{0.01in}\hspace{0.01in}\chi _{\rho \sigma
}(-p)\rangle _{0}=-\left. \frac{i}{p^{2}-m_{\chi }^{2}}\right\vert _{\text{%
PVP}}\left( \frac{\pi _{\mu \rho }\pi _{\nu \sigma }+\pi _{\mu \sigma }\pi
_{\nu \rho }}{2}-\frac{1}{3}\pi _{\mu \nu }\pi _{\rho \sigma }\right)
,\qquad \pi _{\mu \nu }\equiv \eta _{\mu \nu }-\frac{p_{\mu }p_{\nu }}{%
m_{\chi }^{2}},  \label{propp2}
\end{equation}%
where the subscript \textquotedblleft PVP\textquotedblright\ is there to
remind us that $\chi _{\mu \nu }$ must be treated as a PVP inside diagrams.

The projection applies at the classical level as well. The true, classical
theory is neither (\ref{SQG}) nor (\ref{sew}). It is obtained by collecting
the tree diagrams with no $\chi _{\mu \nu }$ external legs and only $\chi
_{\mu \nu }$ internal legs \cite{causalityQG}. The result is%
\begin{equation*}
S_{\text{cl}}(g,\Phi )=-\frac{1}{16\pi G}\int \mathrm{d}^{4}x\sqrt{-g}\left(
2\Lambda +R-\frac{R^{2}}{6m_{\phi }^{2}}\right) +S_{m}(g,\Phi )+\Delta S_{%
\text{nl}}(g,\Phi ),
\end{equation*}%
where $\Delta S_{\text{nl}}$ collects nonlocal vertices that are negligible
at energies lower than $m_{\chi }$.

The prices to pay to have renormalizability and unitarity at the same time
in quantum gravity are the impossibility to distinguish past, present and
future at distances smaller, or intervals shorter, than $1/m_{\chi }$, as
well as a certain \textquotedblleft peak uncertainty\textquotedblright : in
the processes where the PVP is supposed to be \textquotedblleft
detected\textquotedblright , significant complications arise due to the
inherent impossibility of its detection \cite{peak}.

\subsection{Kinetic Lagrangians of Proca and Pauli-Fierz AL-QFTs}

The residue of the propagator (\ref{propp2}) at $p^{2}=m_{\chi }^{2}$ has
the wrong sign. If treated conventionally, it gives a ghost. We know that
the solution, in the realm of local quantum field theory, is to treat it as
a PVP. An alternative option is to alter the propagator (\ref{propp2}) by
means of entire functions, so as to eliminate the zero in the denominator.

The simplest deformation amounts to turning (\ref{propp2}) into%
\begin{equation}
\langle \chi _{\mu \nu }(p)\hspace{0.01in}\hspace{0.01in}\chi _{\rho \sigma
}(-p)\rangle _{0}=\frac{i}{Q(-p^{2}+m_{\chi }^{2})}\left( \frac{\tilde{\pi}%
_{\mu \rho }\tilde{\pi}_{\nu \sigma }+\tilde{\pi}_{\mu \sigma }\tilde{\pi}%
_{\nu \rho }}{2}-\frac{1}{3}\tilde{\pi}_{\mu \nu }\tilde{\pi}_{\rho \sigma
}\right) ,  \label{proppa}
\end{equation}%
where $Q$ is the second option of formula (\ref{Q}) and%
\begin{equation*}
\tilde{\pi}_{\mu \nu }\equiv \eta _{\mu \nu }-\frac{p_{\mu }p_{\nu }}{%
m_{\chi }^{2}}\sigma ^{2}(-p^{2}+m_{\chi }^{2}).
\end{equation*}%
As anticipated, the new \textquotedblleft propagator\textquotedblright\ (\ref%
{proppa}) has no pole, hence it does not actually propagate degrees of
freedom. We can view it as the propagator of a NL-PVP. In the local limit,
the function $\sigma ^{2}$ tends to one, so (\ref{proppa}) tends to (\ref%
{propp2}) (at the tree level), by formula (\ref{limit}).

The choice (\ref{proppa}) corresponds to the nonlocal kinetic Lagrangian%
\begin{eqnarray}
S_{\text{PF}}^{\prime } &=&-\frac{1}{2}\int \mathrm{d}^{4}x\sqrt{-g}\left[
\chi _{\mu \nu }Q(D_{\rho }D^{\rho }+m_{\chi }^{2})\chi ^{\mu \nu }-\chi
Q(D_{\rho }D^{\rho }+m_{\chi }^{2})\chi \right.   \notag \\
&&+2\chi ^{\mu \nu }D_{\mu }\tilde{Q}(D_{\rho }D^{\rho }+m_{\chi
}^{2})D_{\nu }\chi -2\chi ^{\sigma \nu }D_{\mu }\tilde{Q}(D_{\rho }D^{\rho
}+m_{\chi }^{2})D_{\sigma }\chi _{\nu }^{\mu }  \notag \\
&&-\left. R^{\mu \nu }(\chi \chi _{\mu \nu }-2\chi _{\mu \rho }\chi _{\nu
}^{\rho })\right] ,  \label{SpFP}
\end{eqnarray}%
where%
\begin{equation*}
\tilde{Q}(x)=\frac{x}{m_{\chi }^{2}+(x-m_{\chi }^{2})\sigma ^{2}(x)}.
\end{equation*}%
It is not necessary, at this stage, to modify the nonminimal coupling (last
line).

The lagrangian of $S_{\text{PF}}^{\prime }$ is singular for $x=0$ and $%
(m_{\chi }^{2}-x)\sigma ^{2}(x)=m_{\chi }^{2}$ , where $x=m_{\chi
}^{2}-p^{2} $. The singularities are simple poles in $x$, and can be
prescribed by means of the Cauchy principal value. This keeps $S_{\text{PF}%
}^{\prime }$ convergent and real.

For reasons similar to those explained in the case of PVPs, the action $S_{%
\text{PF}}^{\prime }$ is not the true classical action, but a sort of
\textquotedblleft interim\textquotedblright\ action. The field $\chi _{\mu
\nu }$ must be integrated out, so the singularity of the Lagrangian is
harmless. What is important is that the propagator (\ref{proppa}) is
regular. Below we show that the vertices are regular as well.

If we use $\pi _{\mu \nu }$ in (\ref{proppa}), instead of $\tilde{\pi}_{\mu
\nu }$, we have (\ref{SpFP}) with $\tilde{Q}\rightarrow \sigma ^{-2}$. Then
the Lagrangian has singularities $\sim 1/x^{2}$, which are more severe, to
the extent that the action $S_{\text{PF}}^{\prime }$ becomes also singular.
Again, what is important is that the propagator and the vertices are regular.

For reference, let us consider the action%
\begin{equation*}
S_{\text{Proca}}=\frac{1}{4}\int \mathrm{d}^{4}x\sqrt{-g}\left[ g^{\mu \rho
}g^{\nu \sigma }(\partial _{\mu }A_{\nu }-\partial _{\nu }A_{\mu })(\partial
_{\rho }A_{\sigma }-\partial _{\sigma }A_{\rho })-2m^{2}g^{\mu \nu }A_{\mu
}A_{\nu }\right]
\end{equation*}%
of a PVP Proca vector $A_{\mu }$. Using the same notation as above with $%
m_{\chi }\rightarrow m$, the propagator%
\begin{equation*}
\langle A_{\mu }(p)\hspace{0.01in}\hspace{0.01in}A_{\nu }(-p)\rangle
_{0}=\left. \frac{i\pi _{\mu \nu }}{p^{2}-m^{2}}\right\vert _{\text{PVP}}
\end{equation*}%
can be deformed into 
\begin{equation*}
\langle A_{\mu }(p)\hspace{0.01in}\hspace{0.01in}A_{\nu }(-p)\rangle _{0}=-%
\frac{i}{Q(-p^{2}+m^{2})}\tilde{\pi}_{\mu \nu },
\end{equation*}%
which is derived from the modified action%
\begin{equation*}
S_{\text{Proca}}^{\prime }=-\frac{1}{2}\int \mathrm{d}^{4}x\sqrt{-g}\left[
g^{\mu \nu }A_{\mu }Q(D_{\rho }D^{\rho }+m^{2})A_{\nu }+A_{\mu }D^{\nu }%
\tilde{Q}(D_{\rho }D^{\rho }+m^{2})D^{\mu }A_{\nu })\right] .
\end{equation*}

\subsection{AL-QG}

Summarizing, asymptotically local quantum gravity is the theory described by
the action%
\begin{equation}
S_{\text{QG}}(\tilde{g},\phi ,\chi ,\Phi )=\tilde{S}_{\text{HE}}(\tilde{g}%
)+S_{\chi }^{\prime }(\tilde{g},\chi )+S_{\phi }(\tilde{g}+\psi ,\phi
)+S_{m}(\tilde{g}\mathrm{e}^{\kappa _{\phi }\phi }+\psi \mathrm{e}^{\kappa
_{\phi }\phi },\Phi ),  \label{sew2}
\end{equation}%
where%
\begin{equation}
S_{\chi }(g,\chi )=-S_{\text{PF}}^{\prime }(g,\chi ,m_{\chi }^{2})+S_{\chi
}^{(>2)}(g,\chi ).  \label{sc2}
\end{equation}%
It is obtained from (\ref{sew}) by replacing the Pauli-Fierz $\chi _{\mu \nu
}$ action with (\ref{SpFP}).

In fact, (\ref{sew2}) is just the starting action, because, as we have shown
in the previous section, the multi-threshold sectors of involved diagrams
may need to be adjusted along the way, in order to reach the correct local
limit. Moreover, the deformed theory is not guaranteed to be renormalizable,
so further adjustments may be required. We discuss this issue below.

The field $\chi _{\mu \nu }$ does not need a special prescription in AL-QG,
since the nonlocal deformation (\ref{proppa}) of the propagator is
self-sufficient. Modulo the adjustments just mentioned, the local limit of (%
\ref{sc2}) is the theory of quantum gravity with purely virtual particles 
\cite{LWgrav}, that is to say, (\ref{SQG}) or (\ref{sew}), with $\chi _{\mu
\nu }$ treated as a PVP.

\subsection{Vertices}

\label{vertices}

Now we show that the vertices obtained by expanding (\ref{sc2}) around flat
space are entire functions (see \cite{Marcus} and \cite{Lanza}). Write%
\begin{equation*}
D_{\mu }D^{\mu }+m^{2}=\Box +V_{\mu }\partial ^{\mu }+W,
\end{equation*}%
where $\Box $ is the flat-space D'Alembertian. Let $f(z)=\sum_{n=0}^{\infty
}a_{n}z^{n}$ denote a generic entire function. We assume that $f(D_{\mu
}D^{\mu }+m^{2})$ belongs to an expression where the derivatives can be
integrated by parts. Expanding, the first order in $V$ and $W$ is 
\begin{eqnarray}
&&\sum_{n=0}^{\infty }a_{n}\sum_{k=0}^{n-1}\Box ^{k}(V_{\mu }\partial ^{\mu
}+W)\Box ^{n-k-1}=(V_{\mu }\partial ^{\mu }+W)\sum_{n=0}^{\infty
}a_{n}\sum_{k=0}^{n-1}\overleftarrow{\Box }^{k}\Box ^{n-k-1}  \notag \\
&&\qquad =(V_{\mu }\partial ^{\mu }+W)\sum_{n=0}^{\infty }a_{n}\frac{%
\overleftarrow{\Box }^{n}-\Box ^{n}}{\overleftarrow{\Box }-\Box }=(V_{\mu
}\partial ^{\mu }+W)f(\overleftarrow{\Box },\Box ),  \label{incre}
\end{eqnarray}%
where 
\begin{equation}
f(x,y)\equiv \frac{f(x)-f(y)}{x-y}  \label{entire}
\end{equation}%
denotes the incremental ratio of the function $f$, and the arrow on $%
\overleftarrow{\Box }$ means that the box acts to the very left, beyond $%
V_{\mu }\partial ^{\mu }+W$. In momentum space, the ratio is calculated with
respect to the right and left momenta. Formula (\ref{entire}) shows that if $%
f(x)$ is entire, $f(x,y)$ is also entire, that is to say, the incremental
ratio of an entire function is an entire function (of two variables): no
singularity appears. Moreover, with $f=1/Q$ the ratio converges to 
\begin{equation}
\frac{Q^{-1}(-p^{2}+m_{\chi }^{2})-Q^{-1}(-q^{2}+m_{\chi }^{2})}{-p^{2}+q^{2}%
}\rightarrow \mathcal{P}\frac{\frac{1}{p^{2}-m_{\chi }^{2}}-\frac{1}{%
q^{2}-m_{\chi }^{2}}}{p^{2}-q^{2}}=-\mathcal{P}\frac{1}{p^{2}-m_{\chi }^{2}}%
\frac{1}{q^{2}-m_{\chi }^{2}}  \label{PP}
\end{equation}%
in the cone $\mathcal{C}$ fast enough to validate the arguments of the
previous sections. Note that (\ref{PP}) is the correct result for PVPs at
the tree level\footnote{%
It is not correct inside loop diagrams, but we know from section \ref%
{WickDirect} that the loop integrals make the right PVP\ expressions appear
there as well, possibly up to multi-thresholds.}.

To the second order in some operator $\delta $, we find, with $r=\Box $, 
\begin{equation}
f(r+\delta )=\sum_{n=0}^{\infty }a_{n}(r+\delta )^{n}\rightarrow
\sum_{n=0}^{\infty }a_{n}\sum_{k=0}^{n-2}\sum_{l=0}^{n-k-2}r^{k}\delta
r^{l}\delta r^{n-k-l-2}=\delta _{12}\delta _{23}\frac{%
f(r_{1},r_{3})-f(r_{2},r_{3})}{r_{1}-r_{2}},  \label{second}
\end{equation}%
where the subscripts have been introduced to keep track of the ordering of
the various ingredients, which is $r_{1}\delta _{12}r_{2}\delta _{23}r_{3}$.
For example, in momentum space the ordering gives%
\begin{equation*}
r(p+k+q)\delta (k)r(p+q)\delta (q)r(p),
\end{equation*}%
where $k$ and $q$ are the incoming momenta of the insertions $\delta _{12}$
and $\delta _{23}$, respectively.

Formula (\ref{second}) shows that the second-order vertex is the incremental
ratio of the incremental ratio,%
\begin{equation}
f(x,y,z)=\frac{f(x,z)-f(y,z)}{x-y},  \label{entire2}
\end{equation}%
keeping one variable fixed (which one being immaterial). In particular, the
vertex is well defined and tends to its local limit fast enough. One can
proceed similarly for the rest of the expansion.

To derive the loop integrals of the theory (\ref{sc2}), we proceed as
follows. First, we integrate out $\chi _{\mu \nu }$. This generates
\textquotedblleft functional diagrams\textquotedblright\ built by means of
the covariantized versions of the propagators (\ref{proppa}), where $1/Q$
and $\sigma $ are (entire) functions of $D_{\mu }D^{\mu }+m^{2}$. Later, we
expand the metric tensor $g_{\mu \nu }$ around flat space. The expansion
also acts inside the covariantized $\chi _{\mu \nu }$ propagators, and is
treated by means of identities like (\ref{incre}). By the results of this
section, it generates only entire functions, through repeated incremental
ratios. Once the expansion is worked out to the order of our interest, we
are ready to study the loop integrals. They are well defined, because the
vertices are entire functions, the propagators of physical particles are
standard ones and the propagators of PVPs are also entire functions.

Summarizing,

\begin{enumerate}
\item the covariantized Lagrangians of the theories considered here generate
entire functions $1/Q$ and $\sigma $ of the covariant box, upon integrating
out the nonlocal purely virtual fields;

\item the vertices, obtained by expanding the covariant derivatives as
simple derivatives plus corrections proportional to fields, involve repeated
incremental ratios;

\item the repeated incremental ratios of entire functions, formulas (\ref%
{entire}) and (\ref{entire2}), are entire functions of more variables,
because they can be expanded in power series of all their variables at the
same time\footnote{%
Their expressions show that the zeros of the denominators are compensated by
the zeros of the numerators, so no singularity occurs.};

\item because of this, the diagrammatic structure is the one exhibited in
formula (\ref{BM}).
\end{enumerate}

That is to say, the asymptotically local theories we have formulated, such
as the nonlocal deformation (\ref{sc2}) of quantum gravity with purely
virtual particles, lead to Feynman diagrams where the vertices and the
propagators are made of entire functions, apart from the propagators of the
physical particles, which have the usual poles.

\subsection{Arbitrariness and renormalization}

The AL-QG theory (\ref{sc2}) is unitary, and so is its local limit, which
is, by construction, the PVP-QG theory of quantum gravity described by the
action (\ref{sew}), with $\chi _{\mu \nu }$ treated as a PVP. Given that (%
\ref{sew}) is renormalizable, although not manifestly, it is mandatory to
inquire whether AL-QG is also renormalizable or not.

The nonlocal theories of the literature \cite{kuzmin,Tomboulis,Modesto} are
super-renormalizable. Unfortunately, their renormalization properties do not
extend to nonlocal deformations of strictly renormalizable theories, such as
(\ref{SQG})  \cite{Lanza}. In view of this, AL-QG is not expected to be
renormalizable either. For the reasons that we explain below, we do not
think that this is a serious liability.

Although PVP-QG is unique \cite{LWgrav}, the nonlocal extension PVP-QG $%
\rightarrow $ AL-QG is not. Any entire function like those considered in the
literature \cite{kuzmin,Tomboulis,Modesto} can be used for $h$ inside the
second option for the function $Q$ in (\ref{Q}). This leads to an infinite
arbitrariness. The arbitrariness likely turned on by renormalization,
together with the one associated with the adjustments of the multi-threshold
contributions mentioned in the previous section, is not worse than that.

Moreover, the problem of arbitrariness is predicated on the assumption that
the nonlocal theories are fundamental ones, but this is not what we claim
here. We merely view AL-QFTs as tools to move beyond common frameworks in
quantum field theory.

The main weakness of nonlocal quantum field theory is the lack of a
fundamental principle for selecting the form factors that modify the
propagators. In our approach this problem is addressed by the very existence
of the local limit in Minkowski spacetime, in the sense that we can view 
\textit{that} as the missing guiding principle. Then the candidate theories
of the universe are the local limits themselves.

The arbitrariness of AL-QFT is reminiscent of the one of off-the-mass shell
physics \cite{offshell}. In the latter, the extra parameters are not
properties of the fundamental interactions, but describe the environment
where the phenomenon is observed, such as the experimental apparatus and the
observer itself. We could say that they account for the quantum/classical
interplay between the phenomenon and the rest of the universe. It would be
interesting to uncover the map relating the arbitrariness of AL-QFT to the
one of the off-shell approaches to QFT of ref.s \cite{offshell}. We postpone
this task, because it is beyond the scope of this paper.

We stress that the non uniqueness of the nonlocal deformation does not imply
a lack of predictivity. The number of parameters impacting the phenomenon we
want to observe is hopefully finite. Once they are identified, they can be
fixed by sacrificing an equal number of initial measurements, after which
every other measurement is predicted efficiently.

\section{Conclusions}

\label{conclusions}\setcounter{equation}{0}

We have investigated the relationship between nonlocal and local quantum
field theories, and searched for a practical definition of local limit. On
general grounds, it is not obvious that a nonlocal model admits a local
limit of some sort.

The nonlocal deformations are encoded into form factors that multiply the
propagators and remove poles that otherwise would propagate ghosts. We have
shown that the form factors inspired by the models studied in the current
literature give theories that have well-defined limits only in Euclidean
space. Singular behaviors appear in the Minkowskian correlation functions.

To overcome this difficulty, we have relaxed certain requirements and
defined a new class of unitary, asymptotically local\ theories, which have
well-defined local limits in Minkowski spacetime. Unitarity forbids target
models with ghosts and privileges models with purely virtual particles.

Inside the bubble diagram, the nonlocal deformation generates PVPs directly
in the limit. In the triangle diagram, it does so possibly up to
multi-threshold corrections, which can be adjusted by fine tuning the
deformation itself.

The asymptotically local deformation of a local theory is not unique, and
not renormalizable. This is not a liability, in our approach, because we are
not proposing AL-QFTs as candidate fundamental theories of nature, but
merely as tools to provide alternative formulations of theories with PVPs,
approximations to study the violation of microcausality and the peak
uncertainty, or alternative approaches to off-shell physics, more commonly
treated by restricting QFT\ to a finite interval of time and a compact space
manifold.

In off-shell physics, the non uniqueness is associated with the possibility
of introducing parameters that describe the apparatus, the classical
environment surrounding the experiment and the observer itself. At the
practical level, the theory is predictive in the following sense. One first
identifies the parameters that matter to the phenomenon under consideration.
Assuming that they are finitely many-- as is reasonable to expect under
normal circumstances, or\ in cleverly arranged setups --, one sacrifices an
equal number of initial measurements to fix them. After that, one can verify
whether the subsequent measurements are correctly predicted or not.

We believe that the nonlocality-locality relation worked out here can
stimulate further investigations and some out-of-the-box thinking.

\vskip.5truecm \noindent {\large \textbf{Acknowledgments}}

The author is grateful to G. Calcagni, Jiangfan Liu and L. Modesto for
helpful discussions.

\vskip1truecm

{\textbf{\huge Appendices}} \renewcommand{\thesection}{\Alph{section}} %
\renewcommand{\theequation}{\thesection.\arabic{equation}} %
\setcounter{section}{0}

\section{Key functions}

\label{approx}\setcounter{equation}{0}

In this appendix we give entire functions that tend to the absolute value,
the sign function and the Cauchy principal value in the local limit $\lambda
\rightarrow \infty $. They are used in section \ref{AL} to build the models
studied in the paper.

\subsection{Approximating the absolute value}

The absolute value of a complex number $z$ can be approximated by the never
vanishing entire function%
\begin{equation}
h(z)\equiv \exp \left( \frac{1}{2}\int_{0}^{z^{2}}\frac{1-\mathrm{e}^{-w}}{w}%
\mathrm{d}w-\frac{\gamma _{\text{E}}}{2}\right) =\exp \left( \frac{1}{2}\ln
z^{2}+\frac{1}{2}\Gamma (0,z^{2})\right) .  \label{hz}
\end{equation}%
Precisely, in the double cone $\mathcal{C}=\{z:-\pi /4<$ arg[$z$] $<\pi /4$
or $3\pi /4<$ arg[$z$] $<5\pi /4\}$, we have \cite{NIST}%
\begin{equation}
h(z)=\sqrt{z^{2}}\left[ 1+\frac{\mathrm{e}^{-z^{2}}}{2z^{2}}\left( 1+%
\mathcal{O}\left( \frac{\mathrm{e}^{-z^{2}}}{z^{2}}\right) \right) \left( 1+%
\mathcal{O}\left( \frac{1}{z^{2}}\right) \right) \right] ,  \label{expansion}
\end{equation}%
so%
\begin{equation}
\lim_{\lambda \rightarrow +\infty }\frac{h(\lambda z)}{\lambda }=\sqrt{z^{2}}%
\text{\quad for }z\in \mathcal{C},\qquad \lim_{\lambda \rightarrow +\infty }%
\frac{h(\lambda x)}{\lambda }=|x|\quad \text{\ for }x\in \mathbb{R}.
\label{ess}
\end{equation}

It is important to stress that $h(z)$ has no zeros. Its reciprocal $1/h(z)$
is useful to build the \textquotedblleft propagators\textquotedblright\ that
tend to purely virtual particles in the local limit.

\begin{figure}[t]
\begin{center}
\includegraphics[width=6truecm]{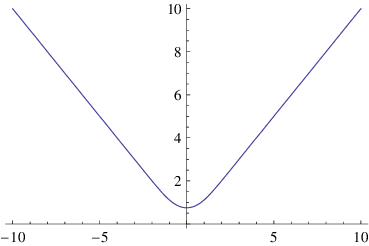}
\end{center}
\caption{Plot of $h(x)$ for real $x$}
\label{h}
\end{figure}

The function $h(z)$ is even,%
\begin{equation}
h(z)=h(-z),  \label{odd}
\end{equation}%
and satisfies 
\begin{equation}
h^{\ast }(z)=h(z^{\ast }).  \label{reality}
\end{equation}%
Moreover, on the real axis it is positive and bounded from below by the
absolute value, as illustrated in fig. \ref{h}:%
\begin{equation}
h(x)\geqslant |x|,\qquad x\in \mathbb{R}\text{.}  \label{bound}
\end{equation}

Finally, (\ref{ess}) gives%
\begin{equation}
\lim_{\lambda \rightarrow +\infty }\frac{h^{2}(\lambda z)}{\lambda ^{2}z}%
=z\quad \text{for }z\in \mathcal{C}.  \label{poly}
\end{equation}

\subsection{Approximating the sign function}

The sign function sgn$(x)$ can be approximated on the complex plane by the
entire function%
\begin{equation}
\sigma (z)=\frac{z}{h(z)},  \label{sigmaz}
\end{equation}%
which vanishes at the origin. In $\mathcal{C}$ the rescaled function $\sigma
(\lambda z)$ tends to $z/\sqrt{z^{2}}$ for $\lambda \rightarrow \infty $,
thus%
\begin{equation}
\lim_{\lambda \rightarrow +\infty }\sigma (\lambda z)=\text{sgn}(\text{Re}%
[z])\quad \text{for }z\in \mathcal{C},\qquad \lim_{\lambda \rightarrow
+\infty }\sigma (\lambda x)=\text{sgn}(x)\quad \text{for }x\in \mathbb{R}.
\label{limk}
\end{equation}

Note that $\sigma (z)$ is odd in the complex plane:%
\begin{equation}
\sigma (z)=-\sigma (-z).  \label{even}
\end{equation}%
Moreover, (\ref{bound}) gives 
\begin{equation}
|\sigma (x)|\leqslant 1,\qquad x\in \mathbb{R}\text{.}  \label{bound2}
\end{equation}

\subsection{Approximating the principal value}

The combination $\sigma (x)/h(x)=x/h^{2}(x)$ approximates the Cauchy
principal value $\mathcal{P}$ on the real axis. Specifically, we prove that%
\begin{equation}
\lim_{\lambda \rightarrow +\infty }\frac{\lambda \sigma (\lambda x)}{%
h(\lambda x)}=\mathcal{P}\frac{1}{x},\qquad x\in \mathbb{R},  \label{limit}
\end{equation}%
in the sense of distributions.

First observe that 
\begin{equation}
\lim_{\lambda \rightarrow +\infty }\frac{\lambda \sigma (\lambda x)}{%
h(\lambda x)}=\frac{1}{x}  \label{almost}
\end{equation}%
for every real $x\neq 0$. This result follows from (\ref{ess}) and (\ref%
{limk}).

If $\varphi (x)$ denotes a real test function, consider the integral%
\begin{equation}
\mathcal{I}\equiv \lim_{\lambda \rightarrow +\infty }\int_{-\infty
}^{+\infty }\mathrm{d}x\frac{\lambda \sigma (\lambda x)\varphi (x)}{%
h(\lambda x)},  \label{I}
\end{equation}%
which can also be written as 
\begin{equation}
\mathcal{I}=\lim_{\lambda \rightarrow +\infty }\int_{-\infty }^{+\infty }%
\mathrm{d}x\frac{\lambda \sigma (\lambda x)(\varphi (x)-\varphi (-x))}{%
2h(\lambda x)},  \notag
\end{equation}%
thanks to (\ref{odd}) and (\ref{even}). The modulus of the $\mathcal{I}$
integrand is bounded above by a $\lambda $-independent function that is
integrable on $\mathbb{R}$. Indeed, the inequalities (\ref{bound}) and (\ref%
{bound2}) give%
\begin{equation}
\left\vert \frac{\lambda \sigma (\lambda x)(\varphi (x)-\varphi (-x))}{%
2h(\lambda x)}\right\vert \leqslant \frac{1}{2}\left\vert \frac{\varphi
(x)-\varphi (-x)}{x}\right\vert ,  \label{a}
\end{equation}%
for real $x$. Then the dominated convergence theorem allows us to exchange
the limit with the integral. Given that (\ref{almost}) holds almost
everywhere, we obtain%
\begin{equation}
\mathcal{I}=\int_{-\infty }^{+\infty }\mathrm{d}x\frac{\varphi (x)-\varphi
(-x)}{2x}=\mathcal{P}\int_{-\infty }^{+\infty }\frac{\mathrm{d}x}{x}\varphi
(x),  \label{theo}
\end{equation}%
as we wished to show.

\section{Bubble diagram with PVPs}

\label{app1}\setcounter{equation}{0}

In this appendix we review the calculation of the bubble diagram with one or
two circulating PVPs, in the formulation of ref. \cite{diagrammarMio}. We
work in Minkowski spacetime, so here $p$ and $k$ denote Minkowskian momenta.

We start from the integral%
\begin{equation*}
\mathcal{B}_{\text{ph}}\equiv \int \frac{\mathrm{d}^{D}p}{(2\pi )^{D}}\frac{1%
}{p^{2}-m^{2}+i\epsilon }\frac{1}{(p-k)^{2}-m^{2}+i\epsilon ^{\prime }}
\end{equation*}%
of the standard bubble with circulating physical particles. First, we
decompose the propagators 
\begin{equation*}
\frac{1}{q^{2}-m^{2}+i\epsilon }\rightarrow \frac{1}{2\omega _{\mathbf{q}}}%
\left( \frac{1}{q^{0}-\omega _{\mathbf{q}}+i\epsilon }-\frac{1}{q^{0}+\omega
_{\mathbf{q}}-i\epsilon }\right) 
\end{equation*}%
by isolating the particle and antiparticle poles, where $\omega _{\mathbf{q}%
}=\sqrt{\mathbf{q}^{2}+m^{2}}$ is the frequency. Then we expand the
integrand and integrate on $p^{0}$ by means of the residue theorem. The
result is%
\begin{equation*}
\mathcal{B}_{\text{ph}}=\int \frac{\mathrm{d}^{D-1}\mathbf{p}}{(2\pi )^{D-1}}%
\frac{i}{4\omega _{\mathbf{p}}\omega _{\mathbf{p}-\mathbf{k}}}\left( \frac{1%
}{k^{0}+\omega _{\mathbf{p}-\mathbf{k}}+\omega _{\mathbf{p}}-i(\epsilon
+\epsilon ^{\prime })}-\frac{1}{k^{0}-\omega _{\mathbf{p}-\mathbf{k}}-\omega
_{\mathbf{p}}+i(\epsilon +\epsilon ^{\prime })}\right) .
\end{equation*}%
At this point, we use%
\begin{equation*}
\frac{1}{x+i\epsilon }=\mathcal{P}\frac{1}{x}-i\pi \delta (x)
\end{equation*}%
for each term inside the parentheses. We obtain 
\begin{eqnarray}
\mathcal{B}_{\text{ph}} &=&\mathcal{P}\int \frac{\mathrm{d}^{D-1}\mathbf{p}}{%
(2\pi )^{D-1}}\frac{i}{4\omega _{\mathbf{p}}\omega _{\mathbf{p}-\mathbf{k}}}%
\left( \frac{1}{k^{0}+\omega _{\mathbf{p}-\mathbf{k}}+\omega _{\mathbf{p}}}-%
\frac{1}{k^{0}-\omega _{\mathbf{p}-\mathbf{k}}-\omega _{\mathbf{p}}}\right) 
\notag \\
&&-\int \frac{\mathrm{d}^{D-1}\mathbf{p}}{(2\pi )^{D-1}}\frac{\pi }{4\omega
_{\mathbf{p}}\omega _{\mathbf{p}-\mathbf{k}}}\left[ \delta (k^{0}+\omega _{%
\mathbf{p}-\mathbf{k}}+\omega _{\mathbf{p}})+\delta (k^{0}-\omega _{\mathbf{p%
}-\mathbf{k}}-\omega _{\mathbf{p}})\right] .  \label{rebub}
\end{eqnarray}

The bubble diagram $\mathcal{B}_{\text{PVP}}$ with one or two circulating
PVPs is obtained from $\mathcal{B}_{\text{ph}}$ by dropping the delta terms:%
\begin{equation}
\mathcal{B}_{\text{PVP}}=\mathcal{P}\int \frac{\mathrm{d}^{D-1}\mathbf{p}}{%
(2\pi )^{D-1}}\frac{i}{4\omega _{\mathbf{p}}\omega _{\mathbf{p}-\mathbf{k}}}%
\left( \frac{1}{k^{0}+\omega _{\mathbf{p}-\mathbf{k}}+\omega _{\mathbf{p}}}-%
\frac{1}{k^{0}-\omega _{\mathbf{p}-\mathbf{k}}-\omega _{\mathbf{p}}}\right) .
\label{BPVP}
\end{equation}

For calculations of more complicated diagrams and the general proof of
unitarity, see \cite{diagrammarMio}.

It is important to stress that if we adopt the principal value we see in (%
\ref{MinkLim}) as an alternative propagator in Feynman diagrams, we obtain a
theory of \textquotedblleft Wheelerons\textquotedblright\ \cite{Rocca},
which propagates ghosts. For example, the bubble diagram gives%
\begin{equation}
\mathcal{P}\int \frac{\mathrm{d}^{D}p}{(2\pi )^{D}}\frac{1}{p^{2}-m^{2}}%
\frac{1}{(p-k)^{2}-m^{2}},  \label{wheeler}
\end{equation}%
which has a nonvanishing, unphysical absorptive part \cite{Wheelerons},
absent in (\ref{BPVP}). The moral of the story is that in a theory of PVPs
the radiative corrections are not given by the usual diagrams with a
different propagator, but by new combinations of diagrams \cite%
{PVP20,diagrammarMio}.

\end{document}